\documentclass[journal]{IEEEtran}  % Comment this line out
\IEEEoverridecommandlockouts                              % This command is only
\usepackage{amssymb,amsmath,graphicx,float}
\usepackage{amsthm}
\usepackage{mathtools}

\newtheorem{definition}{Definition}
\newtheorem{proposition}{Proposition}

\newtheorem{remark}{Remark}
\newtheorem{example}{Example}
\newtheorem{problem}{Problem}

\newcommand{\norm}[1]{\left\lVert#1\right\rVert}

\title{\LARGE \bf
	Temporal Logic Inference for Hybrid System Observation with Spatial and Temporal Uncertainties}
\author{Zhe~Xu,~\IEEEmembership{Member,~IEEE},
	Yi~Deng,~\IEEEmembership{Student Member,~IEEE} and Agung~Julius,~\IEEEmembership{Member,~IEEE}
	\thanks{Zhe~Xu is with the Institute
		for Computational Engineering and Sciences (ICES), University of Texas,
		Austin, Austin, TX, Agung~Julius are with the Department of Electrical, Computer, and Systems Engineering, Rensselaer Polytechnic Institute, Troy, NY, USA, Yi~Deng is with East China Institute of Computing Technology, Shanghai, China. E-mail: zhexu@utexas.edu, dengyi267@gmail.com, juliua2@rpi.edu. }
}

\begin{document}
	
	\maketitle \thispagestyle{empty} \pagestyle{empty}
%	
%	\begin{abstract} 
%		In modern smart buildings modeled as hybrid systems, occupancy detection can be cast as observing the discrete states of a hybrid system using the available discrete and continuous system outputs. In this paper, we present a method to construct observers of the hybrid system to distinguish between different locations of the hybrid system by inferring metric temporal logic (MTL) formulae from the simulated trajectories. We first approximate the system behavior by simulating finitely many trajectories with time-robust tube segments around them. These time-robust tube segments account for both spatial and temporal uncertainties that exist in the hybrid system with initial state variations. The inferred MTL formulae classify different time-robust tube segments and thus can be used for classifying the hybrid system behaviors in a provably correct fashion. We implement our approach on a model of a smart building testbed to distinguish two cases of room occupancy.
%	\end{abstract}
	
	\section{Introduction}
	In modern smart buildings, various continuous states such as the temperature, humidity and discrete states such as the air conditioning states have made the system a hybrid system. In a hybrid system, the continuous state keeps flowing in a location (also called a mode or discrete state) until an event is triggered. Then it jumps to a target location and flows continuously again according to possibly different dynamics. As there are increasingly growing interest in finding ways to accurately determine localized building or room occupancy
	in real time, traditional methods seldom apply to multiple dynamics in a hybrid system. 
	
	Presently, there are mainly two categories of approaches for occupancy detection or estimation (e.g. detecting or estimating the number of people in a room). The first category relies on the learning-based techniques such as 
	decision trees \cite{Hailemariam2011} or support vector regression \cite{Hua2016} to find features of different occupancy states from data gathered from various sensors. The second category relies on the mathematical model of systems as they compare available measurements with information analytically derived from the system model. For hybrid systems, the main challenge of the model-based occupancy detection is due to the difficulty in capturing the combined continuous and discrete measurements. 
	
	In this paper, we propose an approach that utilizes both the learning-based techniques and the model-based methods for hybrid system occupancy detection. We mainly focus on distinguishing between different occupancy states and observing the location of the modeled hybrid system at any time. For the learning aspect, there has been a growing interest in learning (inferring) dense-time temporal logic formulae from system trajectories \cite{Kong2017,zheletter2,zhe2016,Bombara2016,zhe2015,zhe_ijcai2019,Asarin2012,Yan2019swarm,zhe2019ACCinfo,Hoxha2017,zheCDC2019GTL,Jin13}. Such temporal logic formulae have been used as high-level knowledge or specifications in many applications in robotics \cite{Allerton2019,zheACC2019DF,zhe_advisory,MH2019IFAC}, power systems \cite{zheACCstorageControl,zhe2017cascade,zhe_control,zheACC2018wind}, smart buildings \cite{zheCDCprivacy,zhe2019privacy}, agriculture \cite{cubuktepe2020policy}, etc. We infer dense-time temporal logic formulae from the temperature and humidity sensor data as dense-time temporal logics can effectively capture the time-related features in the transient period when people enter a room. In the meantime, we also utilize the model information so that the MTL formula that classifies the finite trajectories we gathered also classifies the infinite trajectories that differ from the simulated trajectories by a small margin in both space and time. In our previous work in \cite{zheCDCprivacy}, we have performed classification for trajectories generated from switched systems, which have spatial uncertainties due to initial state variations. In this paper, we extend the results to hybrid systems and we classify time-robust tube segments around the trajectories so that the inferred MTL formula can classify different system behaviors when both the spatial and the temporal uncertainties exist due to initial state variations in a hybrid system. 

To identify the current location, useful information usually originates from comprehensive discrete and continuous system outputs, yet both are of limited availability. For instance, lack of event observability makes it difficult to determine the subsequent locations, especially for those non-deterministic unobservable events that may take place anywhere without any output discrete signal. In that case, we can resort to the available continuous state measurements for the purpose of location identification. Relevant approaches include designing residual generation schemes~\cite{Bayoudh2008,Arogeti2010,Vento2012}, and analyzing the derivatives of the continuous outputs~\cite{Collins2004}. 
The problem can also be addressed by system abstraction, such as abstracting away the continuous dynamics in exchange for temporal information of the discrete event evolution that helps to track locations~\cite{DiBenedetto2008,Zhao2005,DiBenedetto2011,Deng2015_Verification} 

Another aspect of hybrid state estimation is concerned with continuous state tracking for the underlying multiple modes of continuous dynamical systems. In \cite{Balluchi2002,Alessandri2001}, continuous observers are constructed based on the classical Luenberger's approach, and thus continuous state observability is required. In \cite{Tanwani2013}, the authors propose an observer design for switched systems that does not require the continuous systems to be observable. To that end, \cite{Tanwani2013} presents a characterization of observability over an interval. In \cite{YiCDC}, the authors of the present paper propose a framework for hybrid system state estimation from the perspective of bisimulation theory~\cite{Girard2007}. The framework is based on the robust test idea~\cite{Julius2007}, which extracts the spatial and temporal properties of infinitely many trajectories by a finite number of simulations. 
	
\begin{figure}[H]
		\centering
		\includegraphics[scale=0.3]{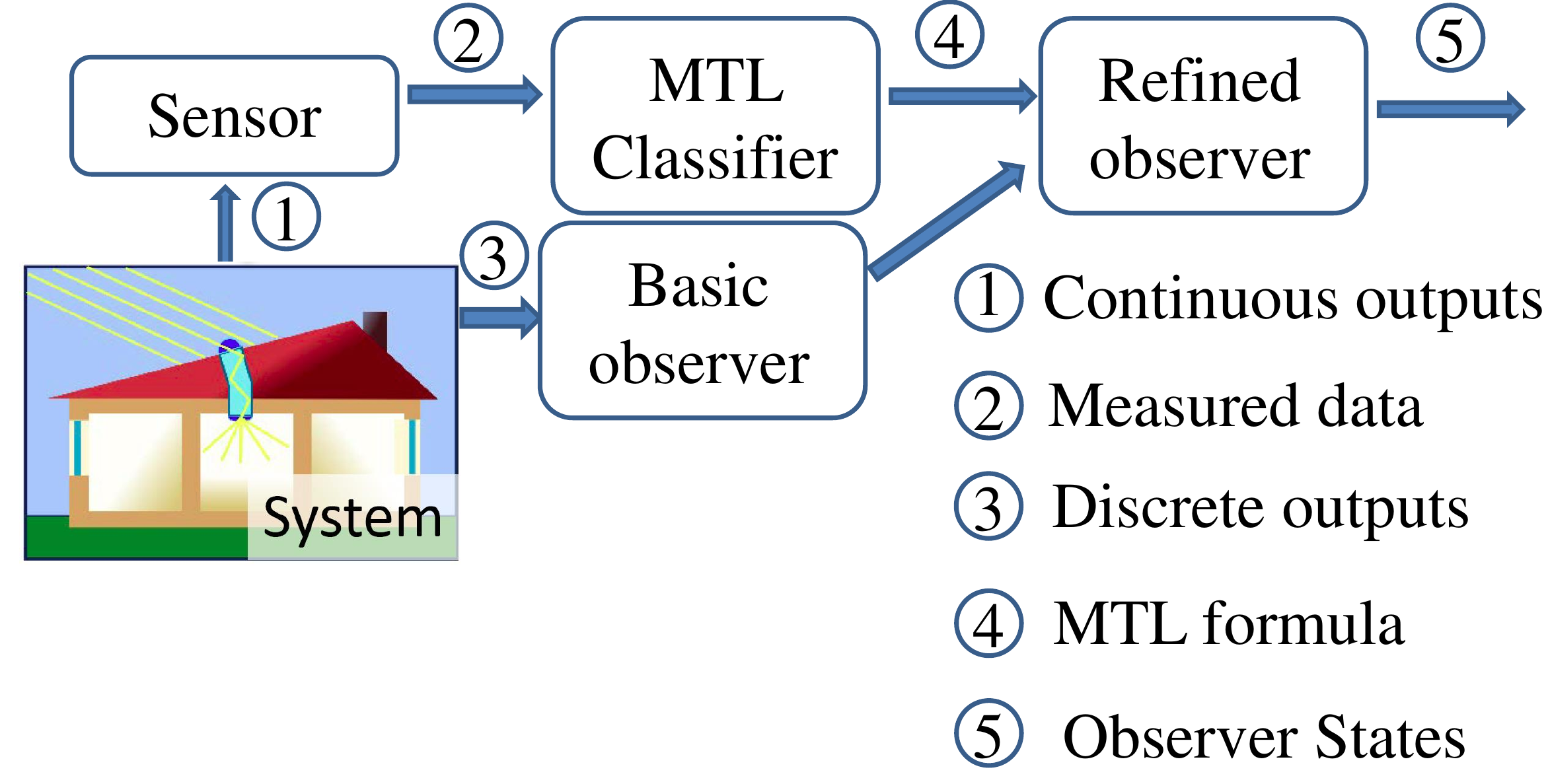}
		\caption{The MTL classifier infers an MTL formula that classifies the sensor data in different conditions (such as different room occupancy states). The refined observer utilizes the MTL formula inferred from the MTL classifier to shrink the basic observer's states.}
		\label{fig_diagapproach}
	\end{figure}

In inferring a temporal logic formula that classifies different system behaviors, we can further design an observer for determining the location of the hybrid system at any time. Our previous work \cite{YiCDC} results in a hybrid observer that estimates both the discrete and continuous states constantly. The observer only uses the discrete outputs generated by the hybrid system's observable events and their timing information as its input, and thus is referred to as the basic observer in this paper. Based on \cite{YiCDC}, we utilize the inferred MTL formula from the MTL classifier to refine the basic observer and the obtained observer is referred to as the refined observer. We illustrate the idea with Figure \ref{fig_diagapproach}. 
	
	\section{Preliminaries}
	\label{sec_hybrid}
	\subsection{Hybrid Automaton}
	A hybrid autonomous system is defined to be a $5$-tuple $\mathcal{H}=(\mathcal{\mathcal{L}}\times \mathcal{\mathcal{X}}, \mathcal{\mathcal{L}}^0\times\mathcal{\mathcal{X}}^0,\mathcal{\mathcal{F}},\mathcal{\mathcal{E}},Inv)$~\cite{Alur1995}:
	\begin{itemize}
		\item $\mathcal{L}\times \mathcal{X}$ is a set of hybrid states $(\ell,x)$, where $\ell\in \mathcal{L}$ is discrete state (location), and $x\in \mathcal{X}$ is continuous state.
		\item $\mathcal{L}^0\times \mathcal{X}^0\subset \mathcal{L}\times \mathcal{X}$ a set of initial states.
		\item $\mathcal{F}=\{f_{\ell}\vert \ell\in \mathcal{L}\}$ associates with each location $\ell\in \mathcal{L}$ the autonomous continuous time-invariant dynamics, $f_\ell: \dot{x}=f_\ell(x)$, which is assumed to admit a unique global solution $\xi_\ell(\tau,x^0_\ell)$, where $\xi_\ell$ satisfies $\frac{\partial\xi_\ell(\tau,x_\ell^0) }{\partial \tau}= f_\ell(\xi_\ell(\tau,x_\ell^0))$, and  $\xi_\ell(0,x^0_\ell)=x_\ell^0$ is the initial condition in $\ell$.% If it is not ambiguous from the context, we will omit the subscript and write $\xi(\tau, x^0)$. 
		\item $Inv:\mathcal{L}\rightarrow 2^\mathcal{X}$ associates an invariant set $Inv(\ell)\subset \mathcal{X}$ with each location. Only if the continuous state satisfies $x\in Inv(\ell)$, can the discrete state be at the location $\ell$.
		\item $\mathcal{E}$ is a set of events. In each location $\ell$, the system state evolves continuously according to $f_\ell$ until an event $e = (\ell, \ell',  g, r), e \in \mathcal{E}$ occurs. The event is guarded by $g \in Inv(\ell)$. Namely, a necessary condition for the occurrence of $e$ is $x \in g$. After the event, the state is reset from $(\ell,x)$ to $(\ell',r(x))$, where $r(x)$ is the reset initial state of $x$.
	\end{itemize}
	When a hybrid system runs, the system state alternately flows continuously and triggers events in $\mathcal{E}$. For convenience, we also define an initialization event $e^0\not\in \mathcal{E}$. Then a trajectory of the system can be defined as a sequence:
	\begin{definition}[Trajectory]
		\label{def_traj}
		A trajectory of a hybrid system H
		is denoted as
		\begin{equation}
			\rho =\{(e^m,\ell^m,x^0_{\ell^m},\tau^m)\}_{m=0}^N,\nonumber
		\end{equation}
		where
		\begin{itemize}
			\item $\forall m\ge 0$, $(\ell^m,x^0_{\ell^m})\in \mathcal{L}\times \mathcal{X}$ are the (reset) initial states;
			\item $\forall m\ge 0$, $\tau^m\in\mathbb{R}_{\ge 0}$ (nonnegative real), and $\forall\tau\in [0,\tau^m]$, $\xi_{\ell^m}(\tau,x^0_{\ell^m})\in Inv(\ell^m)$;
			\item $\forall m\ge 1$, $e^m=(\ell^{m-1},\ell^m,g^m,r^m)$, $\xi_{\ell^{m-1}}(\tau^{m-1},$ $x^0_{\ell^{m-1}})\in g^m$, 
			$x^0_{\ell^m}=r^m(\xi_{\ell^{m-1}}(\tau^{m-1},x^0_{\ell^{m-1}}))$, i.e. $(\ell^m,x^0_{\ell^m})$ is the reset initial state for $(\ell^{m-1},\xi_{\ell^{m-1}}(\tau^{m-1},$ $x^0_{\ell^{m-1}}))$.
		\end{itemize}
	\end{definition}
	Each event $e\in E$ has an output symbol $\psi(e)$ that can be observable or unobservable. An unobservable output symbol $\psi(e)$ is specifically denoted as $\epsilon$.
	
	\subsection{Robust Neighborhood Approach}
	\label{sec_robusttest}
	In this section, we briefly review the robust neighborhood approach \cite{Julius2007}, which is based on the approximated bisimulation theory \cite{Girard2007}. 	
	The robust neighborhood approach \cite{Julius2007} is to compute a neighborhood around a simulated initial state, such that any trajectory initiated from the neighborhood will trigger the same event sequence as the simulated trajectory, and the continuous state always stays inside a neighborhood around the continuous state of the simulated one. 
	\begin{definition}
		\label{def_bisimfunc}
		$\Phi_\ell: Inv(\ell) \times Inv(\ell) \rightarrow \mathbb{R}$ is an autobisimulation function for the dynamics of hybrid system $\mathcal{H}$ at location $\ell$, if for any $x_1,x_2\in Inv(\ell)$,
		\[
		\begin{split}
		\Phi_\ell(x_1,x_2)&\ge0,\\
		\frac{\partial{\Phi_\ell(x_1,x_2)}}{\partial{x_1}}f_\ell(x_1)&+\frac{\partial{\Phi_\ell(x_1,x_2)}}{\partial{x_2}}f_\ell(x_2)\le0.
		\end{split}
		\] 
		\label{bisim_def}
	\end{definition}	
	From Definition \ref{bisim_def}, $\Phi_\ell$ can be used to bound the divergence of continuous state trajectories. If we define the level set
	\begin{equation}
		B_\ell(\gamma_\ell,\xi_\ell(\tau,x^0_\ell))\triangleq\{x\vert \Phi_\ell(x,\xi_\ell(\tau,x^0_\ell))<\gamma_\ell\}.
	\end{equation}
	then we can easily conclude that the value of $\Phi_\ell$ is nondecreasing along any two trajectories of
	the system at location $\ell$, i.e. for any initial state $\tilde{x}^0_\ell\in B_\ell(\gamma_\ell,x^0_\ell)$ and $\tau>0$,
	$\xi_\ell(\tau,\tilde{x}^0_\ell)\in B_\ell(\gamma_\ell,\xi_\ell(\tau,x^0_\ell))$.

	Let $e=(\ell,\ell',g,r)$ be an event triggered by a trajectory initiated from $x^0_\ell$. If we want all the trajectories initiated from within 
	$B_\ell(\gamma_\ell,x^0_\ell)$ to avoid triggering a different event $e'=(\ell,\ell'',g',r')$, then we can let
	\[
	\gamma_\ell \le\inf_{y \in g'}\inf_{\tau\in [0,\bar{\tau}]} \Phi_\ell(\xi_\ell(\tau,x^{0}_\ell),y), 
	\]
	where $\bar{\tau}$ is an upper bound of the time for trajectories initiated from $B_\ell(\gamma_\ell,x^0_\ell)$ to transition out of $\ell$ (for details on methods for estimating $\bar{\tau}$, see \cite{Julius2007}). 
	Then for any $\tilde{x}^0_\ell\in B(\gamma_\ell, x^0_\ell), \tau\in [0,\tau]$, we have that $\xi_\ell(\tau,\tilde{x}^{0}_\ell)$ cannot reach $g'$ and thus trigger $e'$. 
	
	Let $\rho=\{(e^m,\ell^m,x^0_{\ell^m},\tau^m)\}_{m=0}^N$ denote the simulated trajectory. We can compute robust neighborhoods $B_{\ell^m}(\gamma_{\ell^m},x^0_{\ell^m})$ around the (reset) initial continuous states $x^0_{\ell^m}$ of $\rho$ such that the property below holds.
	
	\begin{proposition}
		For any initial state $(\ell^0,\tilde{x}_{\ell^0}^0)\in\{\ell^0\}\times B_{\ell^0}(\gamma_{\ell^0},x_{\ell^0}^0)$ and any trajectory $\tilde{\rho}=\{(e^m,\ell^m,\tilde{x}^0_{\ell^m},\tilde{\tau}^m)\}_{m=0}^{N}$ that triggers the same event sequence with the simulated trajectory $\rho$, there exist $\tau_{lead}^m,\tau_{lag}^m>0~(0\le m\le N-1)$ such that
		\begin{itemize}
			\item for all $0\le m\le N-1$, $\tilde{x}^0_{\ell^m}\in B_{\ell^m}(\gamma_{\ell^m},x^0_{\ell^m})$, $\tilde{\tau}^m$ $\in [\tau^m-\tau_{lead}^m,\tau^m+\tau_{lag}^m]$, and $\Phi_{\ell^m}(\xi_{\ell^m}(t,x^0_{\ell^m}),$ $\xi_{\ell^m}(t,\tilde{x}^0_{\ell^m}))\le\gamma_{\ell^m}$ for all $t\in [0,\tilde{\tau}^m]$; 
			\item $\tilde{x}^0_{\ell^N}\in B_{\ell^N}(\gamma_{\ell^N},x^0_{\ell^N})$, and $\Phi_{\ell^N}(\xi_{\ell^N}(t,x^0_{\ell^N}),\xi_{\ell^N}$ $(t,\tilde{x}^0_{\ell^N}))\le\gamma_{\ell^N}$ for all $t\in [0,\min(\tilde{\tau}^N,\tau^N)]$.
		\end{itemize} 
	\end{proposition}

	We simulate trajectories from the initial set $\mathcal{L}^0\times \mathcal{X}^0$ and perform robust neighborhood computation. We denote $\rho_k=\{(e^m_k,\ell^m_k,x^0_{\ell^m_k},\tau^m_k)\}_{m=0}^{N_k}$ as the $k$th simulated trajectory ($k=1,2,\dots$). The robust neighborhood around the (reset) initial state for the segment $m$ of $\rho_k$ is the following:
	\begin{eqnarray}
		B_{\ell^m_k}(\gamma_{\ell^m_k},x^0_{\ell^m_k})
		&=&\{x\vert \Phi_{\ell^m_k}(x^0_{\ell^m_k},x)<\gamma_{\ell^m_k}\},
	\end{eqnarray}
	where $\Phi_{\ell^m_k}$ is the bisimulation function in location $\ell^m_k$, and $\gamma_{\ell^m_k}$ is the radius of the computed robust neighborhood. The initial set can be covered by the robust neighborhoods around the initial states of the finitely simulated trajectories if:
	\begin{equation}
		\label{eq_coverinit}
		\mathcal{L}^0\times\mathcal{X}^0\subset \bigcup\limits_{k} \{\ell_k^0\}\times B_{\ell^0_k}(\gamma_{\ell^0_k},x^0_{\ell^0_k}).
	\end{equation}	
	
\section{Timed Abstraction and Observer Design}
	\label{sec_time}
\subsection{Timed Abstraction}
\label{sec_abstraction}
Based on the simulated trajectories $\{\rho_k\}_{k=1}^{\hat{K}}$, we construct a timed automaton $T = (Q, Q^0, C, \tilde{E}, \tilde{Inv})$~\cite{Alur1994}.
\begin{definition}[Timed Abstraction]
	$T$ consists of
	\begin{itemize}
		\item The state space is $Q:=\{(k,n)\vert k\in \{1,2,\ldots, \hat{K}\}, n\in \{0,1,\ldots, N_k\}\}$.
		\item The initial set is $Q^0:=\{1,2,\ldots,K\}\times\{0\}$.
		\item The set of clock $C$ is a singleton $\{c\}$.
		\item The events $\tilde{e}\in \tilde{E}$ are defined as $\tilde{e}=(q,q',\tilde{g},\tilde{r})$ such that $\tilde{r}(c)=0$, i.e., the only clock is reset after any event, and one of the following cases should be satisfied: 
		\begin{enumerate}
			\item $q=(k,n)$, where $n<N_k$; $q'=(k,n+1)$; and $\tilde{g}=[\tau_k^n-lead_k^n,\tau_k^n+lag_k^n]$; $\tilde{e}$ is associated with the output symbol of $e_k^{n+1}$;
			\item $q=(k,N_k)$, where $k\in [1,K]$; $q'=(k',0)$, where $k'\in Cover(k)$; and $\tilde{g}=[\tau_k^{N_k},\tau_k^{N_k}]$; $\tilde{e}$ is associated with the unobservable output symbol $\epsilon$;
			\item $q=(k,N_k)$, where $k\in [K+1,\hat{K}]$; $q'=EoS$ (end of simulation); and $\tilde{g}=[\tau_k^{N_k},\tau_k^{N_k}]$; $\tilde{e}$ is associated with the unobservable output symbol $\epsilon$;
			\item $q=(k,n)$; $q'=(k',0)$, where $k'\in Ind^f(k,n,e^f)$ for some $e^f\in Feas^f(k,n)$; and $\tilde{g}=[0,\tau_k^n+lag_k^n]$; $\tilde{e}$ is associated with the output symbol of $e^f$.
		\end{enumerate}
		%For the first case, the event $\tilde{e}$ is associated with the same output symbol as $e_k^{n+1}$; for the second and third cases, $\tilde{e}$ is associated with the unobservable output symbol $\epsilon$; for the forth case, $\tilde{e}$ is associated with the output symbol of the faulty event $e^f$.
		\item The invariant set is $\tilde{Inv}(q):=[0,\tau_k^n+lag_k^n]$ if $n<N_k$, $\tilde{Inv}(q):=[0,\tau_k^n]$ if $n=N_k$, where $q = (k,n)\in Q$.
	\end{itemize}
\end{definition}

\begin{figure}
	\centering
	\includegraphics[scale=0.5]{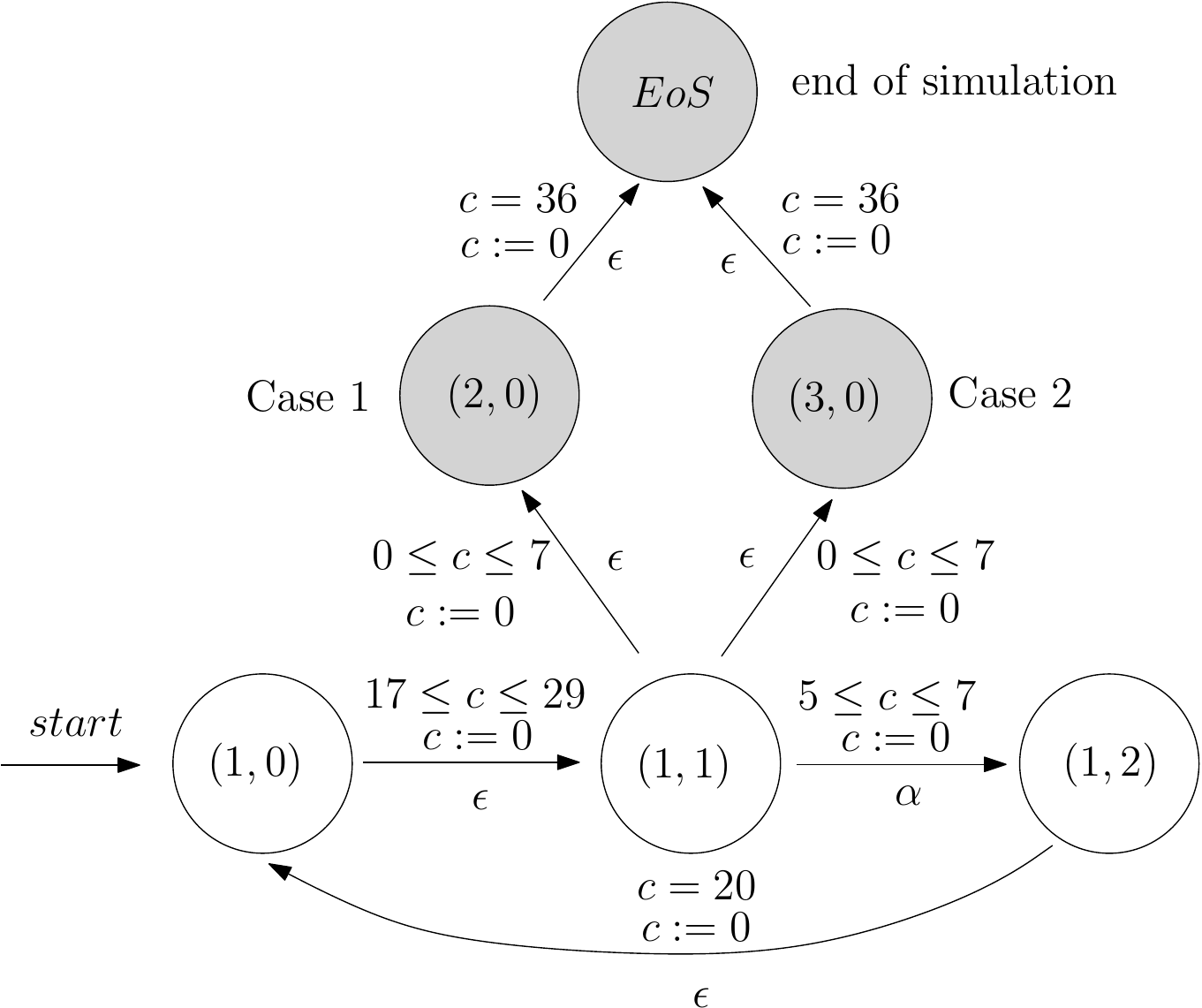}
	\caption{There are one normal trajectory $\rho_1=\{(e^n_1,\ell^n_1,x^n_1,\tau^n_1)\}_{n=0}^2$ and two faulty trajectories $\rho_2 = (e^0_2,\ell^0_2,x^0_2,\tau^0_2)$, $\rho_3 = (e^0_3,\ell^0_3,x^0_3,\tau^0_3)$. The faulty event occurs in $\ell^1_1$ and is unobservable.}
	\label{fig_ta}
\end{figure}

\begin{example}
	Suppose for a system $H$, the following trajectories are simulated in the robust neighborhood approach:
	
	There is one simulated normal trajectory, i.e., $K=1$. The robust neighborhood $Ball(1,0)$ covers the initial set of $H$.
	\[
	\rho_1=(e^0_1,\ell^0_1,x^0_1,\tau^0_1),(e^1_1,\ell^1_1,x^1_1,\tau_1^1),(e^2_1,\ell^2_1,x^2_1,\tau^2_1),
	\]
	where $e^0_1=e^0$ (initialization event), $\tau^0_1=23$, $lead^0_1=lag^0_1=6$; $e^1_1$ is unobservable, $\tau^1_1=6$, $lead^1_1=lag^1_1=1$; $e^2_1$ has the observable output symbol $\alpha$, $\tau^2_1=20$.
	The robust neighborhood $Ball(1,0)$ covers the end of robust tube around $\rho_1$, i.e., $Cover(1)=\{(1,0)\}$. An unobservable faulty event $e^f$ may occur in $\ell^1_1$.
	
	There are two simulated faulty trajectories, i.e., $\hat{K}=3$.
	\[
	\rho_2=(e^0_2,\ell^0_2,x^0_2,\tau^0_2),\quad
	\rho_3=(e^0_3,\ell^0_3,x^0_3,\tau^0_3),
	\]
	where $e^0_2=e^0_3=e^f\in Feas^f(\ell^1_1)$, $\tau_2^0=\tau_3^0=36$. The union of inverse images of the robust neighborhoods $Ball(2,0), Ball(3,0)$ covers the robust tube $Tube(1,1,[0,7])$, i.e., $Ind^f(1,1,e^f)=\{2,3\}$. 
	
	The timed abstraction $T$ is constructed as Fig. \ref{fig_ta}.
\end{example}

Since timed automata can be considered as a subclass of hybrid automata, trajectories and projected timed output symbol sequences can be defined the same way as before. By construction, for any normal trajectory $\rho$ of $H$, there is a trajectory $\tilde{\rho}$ of $T$ such that $\Pi(S(\rho))=\Pi(S(\tilde{\rho}))$. For faulty trajectories, a similar property holds for finite horizon.

\subsection{Basic Observer}
\label{sec_observer}
Based on the timed abstraction $T$, we construct for $H$ an observer $O$. By using the history of system output, i.e., a projected timed output symbol sequence, $O$ over-approximates the set of states reached by $H$. 
%Each state $s$ of $O$ can be represented by a subset of the set
%\begin{eqnarray*}
%\{(k,n)[\bar{a},\bar{b}] &\vert& k,n,\bar{a},\bar{b}\text{ are integers, }\\
% & & 1\le k\le\hat{K}, 0\le n\le N_k,\\
% & & [\bar{a},\bar{b}]\subset \tilde{Inv}((k,n))\}.
%\end{eqnarray*}
%The state of the observer being just updated to $s$ means that the system $H$ must be at some state within $\bigcup_{(k,n)[\bar{a},\bar{b}]\in s} Tube(k,n,[\bar{a},\bar{b}]\cap\mathbb{R}_{\ge 0})$. 

The following definitions are used in the construction of $O$. We illustrate them later with examples.

\begin{definition}
	\label{def_closure}
	Given $T = (Q, Q^0, C, \tilde{E}, \tilde{Inv})$, for each $(k,n)\in Q$, let $Feas: Q\rightarrow 2^{\tilde{E}}$ be the feasible event function:
	\begin{equation}
	Feas((k,n)):=\{\tilde{e}\in\tilde{E}\vert \tilde{e}=((k,n),(k',n'),\tilde{g},\tilde{r})\}.
	\end{equation}
	Let $\bar{a}$ be an integer. Define the $\epsilon[0]$-successors and $\epsilon[0]$-closure of $(k,n)[\bar{a},0]$:
	\begin{eqnarray}
	& & Succ^{\epsilon[0]}((k,n)[\bar{a},0]))\nonumber\\
	&:=&\{(k',n')[\bar{a}-\tilde{b},0] \vert \exists \tilde{e}=((k,n),(k',n'),[0,\tilde{b}],\tilde{r})\nonumber\\
	& & \in Feas((k,n)), \tilde{e}\text{ outputs }\epsilon\}.\\
	& & Cl^{\epsilon[0]}((k,n)[\bar{a},0]))\nonumber\\
	&:=&\{(k,n)[\bar{a},0]\}\cup\{Succ^{\epsilon[0]}((k,n)[\bar{a}',0])\vert\nonumber\\
	& & (k,n)[\bar{a}',0]\in Cl^{\epsilon[0]}((k,n)[\bar{a},0])\}.
	\end{eqnarray}
	Given integers $\bar{a}, \bar{b}$, and $s$ being a set of $(k,n)[\bar{a},\bar{b}]$, define the $\epsilon[0]$-closure of $s$:
	\begin{equation}
	Cl^{\epsilon[0]}(s):=s\cup\{Cl^{\epsilon[0]}((k,n)[\bar{a},0])\vert (k,n)[\bar{a},0]\in s\}.
	\end{equation}
\end{definition}

\begin{figure}
	\centering
	\includegraphics[scale=0.25]{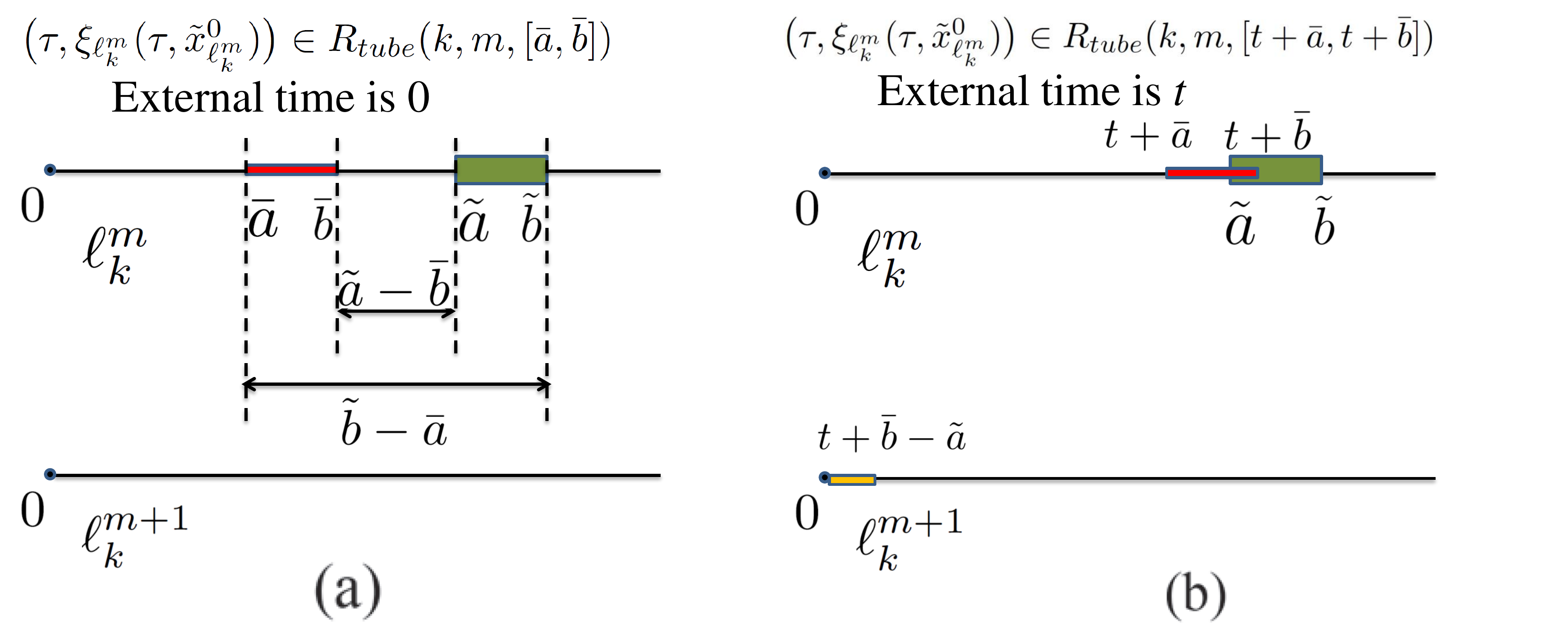}
	\caption{The state $(k,m)[\bar{a}, \bar{b}]$ of the basic observer and an unobservable event that occurs in the time interval $[\tilde{a}, \tilde{b}]$.} 
	\label{update}
\end{figure}

In what follows, we construct an observer that over-approximates the state reached by $H$. Each state of the observer can be represented by a subset of the set
\begin{eqnarray*}
	\{(k,n)[\bar{a},\bar{b}] &\vert& k,n,\bar{a},\bar{b}\text{ are integers, }\\
	& & 1\le k\le\hat{K}, 0\le n\le N_k,\\
	& & [\bar{a},\bar{b}]\subset \tilde{Inv}((k,n))\}.
\end{eqnarray*}
The state of the observer being just updated to $s$ means that the system $H$ must be at some state within $\bigcup_{(k,n)[\bar{a},\bar{b}]\in s} Tube(k,n,[\bar{a},\bar{b}])$. In short, we say the state of $H$ is within $(k,n)[\bar{a},\bar{b}]$ instead of $Tube(k,n,[\bar{a},\bar{b}])$.

For convenience, given $T = (Q, Q^0, C, \tilde{E}, \tilde{Inv})$, we define the feasible event function $Feas: Q\rightarrow 2^{\tilde{E}}$, 
\[
Feas(q):=\{\tilde{e}\in\tilde{E}\vert \tilde{e}=(q,q',\tilde{g},\tilde{r})\}
\]
for each $q\in Q$, and $Feas_{lab}(q)$ as the labels of the feasible events of $q$:
\begin{eqnarray*}
	Feas_{lab}(q):=\{\psi[a,b] &\vert& \exists\tilde{e}=(q,q',\tilde{g},\tilde{r})\in Feas(q),\\
	&& \tilde{e}\rightarrow\psi, \tilde{g}=[a,b]\},
	%In words, $Feas_{lab}(q)$ denote the set of $\psi[a,b]$ such that $\psi$ is the output symbol associated with any $\tilde{e}\in Feas(q)$, and $\tilde{g}=[a,b]$ is the guard of $\tilde{e}$.
\end{eqnarray*}
where $\rightarrow$ means "outputs the symbol".

Suppose the current state $s$ of the observer only contains $\bar{q}=(k,n)[\bar{a},\bar{b}]$, and $s$ is reached at the time instant $t$.  The observer will update the next state based on what is observed after $t$, disregarding what already happened before or at $t$. Consider the feasible events $Feas(q)$ of $q=(k,n)$. The following facts are obvious for a feasible event $\tilde{e}=(q,q',\tilde{g},\tilde{r})\in Feas(q)$ that occurs. Let the label of $\tilde{e}$ be $\psi[a,b]\in Feas_{lab}(q)$.

\begin{itemize}
	\item If $\psi=\epsilon, \bar{b}<a$, then $(a-\bar{b})$ time units later without observation of output symbols, the state of $H$ may be within $q'[0,0]$ or $q[\bar{a}+a-\bar{b},a]$. 
	
	More generally, given any $\tau\in [a-\bar{b},b-\bar{a}]$, $\tau$ time units later, the state of $H$ may be within $q'[0,\tau-(a-\bar{b})]\cap\tilde{Inv}(q')$ or $q[\bar{a}+\tau,\bar{b}+\tau]\cap\tilde{Inv}(q)$. 
	
	We can computationally equivalently let the state estimation at $\tau=a-\bar{b}$ time units later be $q'[-(b-\bar{a}-a+\bar{b}),0]\cup q[\bar{a}+a-\bar{b},a]$ rather than $q'[0,0]\cup q[\bar{a}+a-\bar{b},a]$. These negatively timed states are considered as "latent" reset initial states. The state of $H$ cannot actually be at these negative timed states, which should be kept in mind when interpreting the state read of the observer. Using the latent reset initial states helps us avoid consideration of the unobservable event $\tilde{e}$ triggered at $\tau\in (a-\bar{b},b-\bar{a}]$ time units later.
	
	\item If $\psi\in\Psi_v, \bar{a}<b$, then one should anticipate an observation of $\psi$ at $\tau$ time units later with $\tau\in [\max\{0,a-\bar{b}\},b-\bar{a}]\setminus\{0\}$; after the observation the state of $H$ must be within $q'[0,0]$.
	\item Moreover, let $\tilde{Inv}(q)=[\tilde{a},\tilde{b}]$, $(\tilde{b}-\bar{a})$ time units later, the state of $H$ must be within $q[\tilde{b},\bar{b}+\tilde{b}-\bar{a}]$, or somewhere else through a transition; but it cannot be at $q[\tilde{b},\bar{b}+\tilde{b}-\bar{a}]$ due to the invariant set, so the latter case holds.
\end{itemize}

Given the current observer state $s\ni\bar{q}=(k,n)[\bar{a},\bar{b}]$, and one of the feasible events $\tilde{e}\in Feas(q)$ for $q=(k,n)$, we let the function $Blank(\bar{q},\tilde{e})$ or $Blank^{\tilde{Inv}}(\bar{q})$ (see explanations below) return the shortest blank interval (i.e., no output symbol observed) that the observer needs to wait until it updates the state in order to keep track of the location change of $H$. In specific, corresponding to the three cases listed above, let
\begin{itemize}
	\item $Blank(\bar{q},\tilde{e})=a-\bar{b}$, if $\psi=\epsilon, \bar{b}<a$, since a new location might be visited by $H$ through $\tilde{e}$ and needs to be incorporated into the state of the observer;
	\item $Blank(\bar{q},\tilde{e})=b-\bar{a}$, if $\psi\in\Psi_v, \bar{a}<b$, since up to a blank interval of length $(b-\bar{a})$, the observer is able to determine the absence of $\tilde{e}$. %and update its state accordingly;
	\item $Blank^{\tilde{Inv}}(\bar{q})=\tilde{b}-\bar{a}$, where $\tilde{Inv}(q)=[\tilde{a},\tilde{b}]$, since the state of $H$ has already transitioned out of the location $\ell_k^n$, which requires an update of the observer state.
\end{itemize}

From the discussions above, the state of the observer can be updated from $s$ to $s'$ based on the observation of a blank interval $(t,t']$, where $t$ is the time instant of reaching $s$, and $t'$ is generated through $Blank(\bar{q},\tilde{e})$ or $Blank^{\tilde{Inv}}(\bar{q})$, and becomes the time instant of updating $s'$. The observer can also updates its state to $s''$ by observing $\psi\in\Psi_v$ at some $t''\in (t,t']$. Thus, two types of transition labels should be modeled:
when counting from $t=0$,
\begin{enumerate}
	\item$\epsilon[a_1], a_1>0$, meaning that no symbol is observed during $(0,a_1]$;
	\item $\psi\langle a_2,b], \psi\in\Psi_v,a_2<b$, meaning that $\psi$ is observed during $[a_2,b]$ (or instead, $(a_2,b]$, if $a_2=0$),
\end{enumerate}
Besides these two typical transitions, there are additional types of state updates that have nothing to do with an interval $(t,t']$. Let $t,t',t'',\ldots$ be the time instants when the observer updates states. Clearly, $(t,t'], (t',t''],\ldots$ are either a blank interval, or with an observable event from $H$ at the right end of interval. %and a blank interval may have an unobservable event at the right end as well. 
However, such traces of the observer cannot model the case where multiple events accumulate at the same time instant, that is, an event occurs at an (reset) initial state. So we incorporate two more types of transitions, labeled by $\epsilon[0]$ and $\psi_1\cdots\psi_n\langle a,b],\psi_i\in\Psi_v$.

\begin{definition}
	For $q[\bar{a},0]$, if there exists $\tilde{e}\in Feas(q)$, $\tilde{e}=(q,q',[0,\tilde{b}],\tilde{r})\rightarrow\epsilon$ (meaning $\tilde{e}$ outputs $\epsilon$), then we define $\epsilon[0]$-successors of $q[\bar{a},0]$ as the set of all $q'[-(\tilde{b}-\bar{a}),0]$:
	\begin{eqnarray}
	Succ^{\epsilon[0]}(q[\bar{a},0]) &:=&\nonumber\\
	\{q'[\bar{a}-\tilde{b},0] &\vert& \exists\tilde{e}=(q,q',[0,\tilde{b}],\tilde{r})\in Feas(q)\nonumber\\
	& & \tilde{e}\rightarrow\epsilon\}.
	\end{eqnarray}
\end{definition}
%then include $\epsilon[0]$ or $\psi[0,0]$ in $Feas^{obs}_{lab}(s)$ respectively for the case $\psi=\epsilon, \psi\in\Psi_v$. $Feas^{obs}_{lab}(s)$ is called the feasible transition labels of the observer state $s$. $Feas^{obs}_{lab}(s)$ may also include other labels of the previously mentioned forms $\epsilon[a_1], \psi\langle a_2,b]$ ($a_1>0, \psi\in\Psi_v$) to be defined later. 

The $\epsilon[0]$-extension of $q[\bar{a},0]$, $Ext^{\epsilon[0]}(q[\bar{a},0])$, is defined to be union of $q[\bar{a},0]$, the $\epsilon[0]$-successors of $q[\bar{a},0]$, and the $\epsilon[0]$-successors of all the $\epsilon[0]$-successors.

The transition $\epsilon[0]$ is actually "unobservable" for the observer, since the state transition of $H$ from within $q[\bar{a},0]$ to its $\epsilon[0]$-successor cannot be seen by the observer (the observer can only see observable output symbols and blank intervals), while it indeed causes the state estimation (of $H$ by the observer) changes. So we identify $q[\bar{a},0]$ with $Ext^{\epsilon[0]}(q[\bar{a},0])$ to build an observer state.
Let $s$ be a set of $q[\bar{a},\bar{b}]$, we write
\begin{equation}
Ext^{\epsilon[0]}(s):=s\cup\{Ext^{\epsilon[0]}(q[\bar{a},0])\vert q[\bar{a},0]\in s\}.
\end{equation}

\begin{definition}
	We define the $\psi[0,0]$-successors for $q[0,0]$:
	\begin{eqnarray}
	Succ^{\psi[0,0]}(q[0,0]) &:=&\nonumber\\
	\{q'[0,0] &\vert& \exists \tilde{e}=(q,q',[0,\tilde{b}],\tilde{r})\in Feas(q)\nonumber\\
	& & \tilde{e}\rightarrow\psi\in\Psi_v\cup\{\epsilon\}\}.
	\end{eqnarray}
\end{definition}
%and $\epsilon[0,0]$-successors for $q[0,0]$:
%\begin{equation}
%Su^{\epsilon[0,0]}(q[0,0]) := \{q'[0,0]\vert Feas(q)\ni (q,q',[0,\tilde{b}],\tilde{r})\rightarrow\epsilon\}.
%\end{equation}
%
%The $\epsilon[0,0]$-extension of $q[0,0]$, $Ext^{\epsilon[0,0]}(q[0,0])$, is defined to be union of $q[0,0]$, the $\epsilon[0,0]$-successors of $q[0,0]$, and the $\epsilon[0,0]$-successors of all the $\epsilon[0,0]$-successors.
%
%Let $s$ be a set of $q[0,0]$, define $Ext^{\epsilon[0,0]}(s):=\{Ext^{\epsilon[0,0]}(q[0,0])\vert q[0,0]\in s\}$, $Su^{\psi[0,0]}(s):=\{Su^{\psi[0,0]}(q[0,0])\vert q[0,0]\in s\}$.

Given $q^0[0,0]$, $q^n[0,0]$ is called an extended $\psi[0,0]$-successor of $q[0,0]$ if and only if there exist $q^1[0,0],q^2[0,0],\ldots,q^{n-1}[0,0]$ such that
\begin{equation}
q^i[0,0]\in Succ^{\psi[0,0]}(q^{i-1}[0,0]), i\in\{1,\ldots,n\}.
\end{equation}

Let $\tilde{e}^i\rightarrow\psi^i$ be the event that leads to $q^i$, the concatenation $\psi^1\cdots\psi^n$ is projected to $Proj(\psi^1\cdots\psi^n)$ by erasing all $\psi^i=\epsilon$. If $Proj(\psi^1\cdots\psi^n)\not\in\Psi_v$, then it is considered as a new observable output symbol.

We write $q^0[0,0]\xrightarrow{\psi_{proj}}q^n[0,0]$ to mean that $q^n[0,0]$ is an extended $\psi[0,0]$-successor of $q^0[0,0]$, and the concatenation of output symbols is projected to $Proj(\psi^1\ldots\psi^n)=\psi_{proj}$. The set of extended $\psi[0,0]$-successors of $q[0,0]$ is denoted by $Succ^{\psi[0,0]}_{ext}(q[0,0])$.

\begin{definition}[Basic Observer]
	We construct an basic observer $O=(S,s^0,\bar{\Sigma},f)$ by the following steps, where $S,S^0,\bar{\Sigma},f$ are respectively the state space, initial state, transition labels and transition function:
	\begin{enumerate}
		\item Define $s^0:=Ext^{\epsilon[0]}(\{(1,0)[0,0],\ldots,(K,0)[0,0])\})$. Set $S=\{s^0\}$.
		\item For each new state $s\in S$, compute 
		\begin{eqnarray*}
			& & Blank_{min}(s)\\
			&:=&
			\min\{
			\min\limits_{\bar{q}\in s, \tilde{e}\in Feas(q)}Blank(\bar{q},\tilde{e}), Blank^{\tilde{Inv}}(\bar{q})
			\},
		\end{eqnarray*} 
		where $\bar{q}=q[\bar{a},\bar{b}], q=(k,n)$.
		
		Add the transition label $\epsilon[a], a:=Blank_{min}(s)>0$ to $Feas^{obs}_{lab}(s)$.
		
		Define
		\begin{eqnarray*}
			f'(s,\epsilon[a]) &:=& \{q'[\bar{a}+a-\tilde{b},0]\\
			&\vert & q[\bar{a},\bar{b}]\in s,\\
			& & \tilde{e}=(q,q',[\tilde{a},\tilde{b}],\tilde{r})\in Feas(q),\\
			& & \tilde{e}\rightarrow\epsilon, \bar{b}<\tilde{a}=\bar{b}+a\}.\\
			f(s,\epsilon[a]) &:=& \{q[[\bar{a}+a,\bar{b}+a]\cap\tilde{Inv}(q)]\\
			&\vert& q[\bar{a},\bar{b}]\in s,\\
			& & (\bar{a}+a,\bar{b}+a]\cap\tilde{Inv}(q)\neq\emptyset\}\\
			&\cup& Ext^{\epsilon[0]}(f'(s,\epsilon[a]))
			%\cup\\&  & \{q'[\max\{0,\bar{a}+a-\tilde{b}\},a]\vert q[\bar{a},\bar{b}]\in s, Feas(q)\ni (q,q',[\tilde{a},\tilde{b}],\tilde{r})\rightarrow\epsilon,\\
			%& & \bar{b}\ge\tilde{a},\bar{a}<\tilde{b}\};
		\end{eqnarray*}
		
		Add a transition label $Proj(\psi^1\cdots\psi^n)[a,a], a:=Blank_{min}(s)>0$ into $Feas_{lab}^{obs}(q)$, as long as there exist $q'[\bar{a}',0]\in f'(s,\epsilon[a])$, and $q''[0,0]\in Succ^{\psi[0,0]}_{ext}(q'[0,0])$ through the concatenated output symbols $\psi^1\cdots\psi^n$, and also $Proj(\psi^1\cdots\psi^n)\neq\epsilon$.
		%$Succ^{\psi[0,0]}_{ext}(q'[\bar{a}',0])\neq\emptyset$, letting $\tilde{e}^2\cdots\tilde{e}^n$ be a concatenation of observable events that is used to define an extended $\psi[0,0]$-successor of $q'[\bar{a}',0]$, add all $Trans_{lab}(q'[\bar{a}',0],\tilde{e}^2\cdots\tilde{e}^n)$ in $Feas_{lab}^{obs}(q)$. We have $Trans_{lab}(q'[\bar{a}',0],\tilde{e}^2\cdots\tilde{e}^n)=\psi^2\cdots\psi^n[0,-\bar{a}']$ by definition. Define 
		\begin{eqnarray*}
			f(s, \psi_{proj}[a,a])) &:=& Ext^{\epsilon[0]}(\{q''[0,0]\\
			&\vert& q'[\bar{a}',0]\in f'(s,\epsilon[a]),\\
			& & q'[0,0]\xrightarrow{\psi_{proj}}q''[0,0]\}).
		\end{eqnarray*}
		
		\item Check if there exist $\bar{q}=q[\bar{a},\bar{b}]\in s, q=(k,n)$, $\tilde{e}=(q,q',[\tilde{a},\tilde{b}],\tilde{r})\in Feas(q), \tilde{e}\rightarrow\psi\in\Psi_v$, such that the anticipated interval for the observation of $\psi$, $[\max\{0,\tilde{a}-\bar{b}\},\tilde{b}-\bar{a}]\setminus\{0\}$,  satisfies
		\begin{equation*}
		([\max\{0,\tilde{a}-\bar{b}\},\tilde{b}-\bar{a}]\setminus\{0\})\cap (0,Blank_{min}(s)]\neq\emptyset.
		\end{equation*}
		If so, define $\bar{\sigma}':= \psi\langle\max\{0,\tilde{a}-\bar{b}\}, Blank_{min}(s)]$ for each $\tilde{e}$ that meets the above conditions ($\langle a,b]$ stands for $(a,b]$ if $a=0$, $[a,b]$ if $a>0$). 
		
		Classify the obtained $\bar{\sigma}'$ according to distinct $\psi$. For each classification $[\bar{\sigma}']_{\psi}=\{\psi\langle a_1,b],\psi\langle a_2,b],\ldots\}$, where $b= Blank_{min}(s)$, order the distinct $a_i$ values increasingly and let the result be $a_{(1)}<\ldots<a_{(m)}$.  
		
		Then add to $Feas^{obs}_{lab}(s)$ the transition labels $\{\psi\langle a_{(1)},a_{(2)}), \ldots, \psi\langle a_{(m-1)},a_{(m)}), \psi\langle a_{(m)},b]\}$. 
		
		Obviously, when $m=1$, there is only $\psi\langle a_{(1)},b]$; when $m>1$, these labels are $\{\psi\langle a_{(1)},a_{(2)}), \ldots, \psi[a_{(m-1)},a_{(m)}), \psi[a_{(m)},b]\}$. 
		
		The labels of the form $\psi\langle a,b)$ are a variant of the previously mentioned type $\psi\langle a,b]$. We use them to partition a long interval to short intervals, and make the basic observer a deterministic automaton.
		
		For $\bar{\sigma}=\psi\langle a,b]$ or $\psi\langle a,b)$, $\psi\in\Psi_v$, define
		\begin{eqnarray*}
			f'(s,\bar{\sigma}) &:=&\{q'[0,0]\\
			&\vert& q[\bar{a},\bar{b}]\in s, \tilde{e}=(q,q',[\tilde{a},\tilde{b}],\tilde{r})\\
			& & \tilde{e}\in Feas(q), \tilde{e}\rightarrow\psi,\\
			& & \tilde{b}-\bar{a}\ge b,\tilde{a}-\bar{b}\le a\}.\\
			f(s,\bar{\sigma}) &:=& Ext^{\epsilon[0]}(f'(s,\bar{\sigma})).
		\end{eqnarray*}
		
		Add into $Feas_{lab}^{obs}(q)$ the label $\psi Proj(\psi^1\cdots\psi^n)\langle a,b]$ (or $\psi Proj(\psi^1\cdots\psi^n)\langle a,b)$) as long as there exist $\bar{\sigma}=\psi\langle a,b]$ (or $\psi\langle a,b)$), $q'[0,0]\in f'(s,\bar{\sigma})$, $q''[0,0]\in Succ^{\psi[0,0]}_{ext}(q'[0,0])$ through the concatenated output symbols $\psi^1\cdots\psi^n$ and $Proj(\psi^1\cdots\psi^n)\neq\epsilon$.
		\begin{eqnarray*}
			f(s, \psi\psi_{proj}\langle a,b]) &:=& Ext^{\epsilon[0]}(\{q''[0,0]\\
			&\vert& q'[0,0]\in f'(s,\psi\langle a,b]),\\
			& & q'[0,0]\xrightarrow{\psi_{proj}}q''[0,0]\});
		\end{eqnarray*}
		similar for $\psi\psi_{proj}\langle a,b)$.
		
		\item Include the transition labels $Feas_{lab}^{obs}(s)$ in $\bar{\Sigma}$.
		\item If $s':=f(s,\bar{\sigma})\not\in S$, add the new state $s'$ to $S$.
		\item Repeat Steps 2-5 until no new states are created.
	\end{enumerate}
\end{definition}

Consider the previous example, whose timed abstraction is shown in Fig. \ref{fig_ta}. The basic observer is built by steps below.
\begin{enumerate}
	\item $s^0:=\{(1,0)[0,0]\}$.
	\item $Blank_{min}(s^0)=17$, $Feas_{lab}^{obs}(s^0)=\{\epsilon[17]\}$,\\
	$f'(s^0,\epsilon[17])=\{(1,1)[-12,0]\}$,\\
	$f(s^0,\epsilon[17])=\{[1,0][17,17]\}\cup Ext^{\epsilon[0]}(f'(s^0,\epsilon[17]))$, \\
	$s^1:= \{[1,0][17,17],(1,1)[-12,0],\\ (2,0)[-19,0],(3,0)[-19,0]\}$.
	\item $Blank_{min}(s^1)=\min\{12,19,55,55\}=12$,\\
	$Feas_{lab}^{obs}(s^1)=\{\epsilon[12],\alpha[5,12]\}$,\\
	$f(s^1,\epsilon[12])=\{(1,1)[0,7],(2,0)[-7,0],(3,0)[-7,0]\}$.\\
	$s^2:=\{(1,1)[0,7],(2,0)[-7,0],(3,0)[-7,0]\}$.\\
	$f(s^1,\alpha[5,12])=\{(1,2)[0,0]\}$.\\
	$s^3:=\{(1,2)[0,0]\}$.
	\item $Blank_{min}(s^2)=\min\{7,43,43\}=9$,\\
	$Feas_{lab}^{obs}(s^2)=\{\epsilon[7],\alpha(0,7]\}$,\\
	$f(s^2,\epsilon[7])=\{(2,0)[0,7],(3,0)[0,7]\}$.\\
	$s^4:=\{(2,0)[0,7],(3,0)[0,7]\}$.\\
	$f(s^2,\alpha(5,7])=\{(1,2)[0,0]\}=s^3$.
	\item $Blank_{min}(s^4)=\min\{29,29\}=29$,\\
	$Feas_{lab}^{obs}(s^4)=\{\epsilon[29]\}$,\\
	$f(s^4,\epsilon[29])=\{(3,0)[0,36],(2,0)(3,36)\}$.\\
	$s^5:=\{(3,0)[0,36],(2,0)(3,36)\}$. 
\end{enumerate}

\begin{figure}
	\centering
	\includegraphics[scale=0.36]{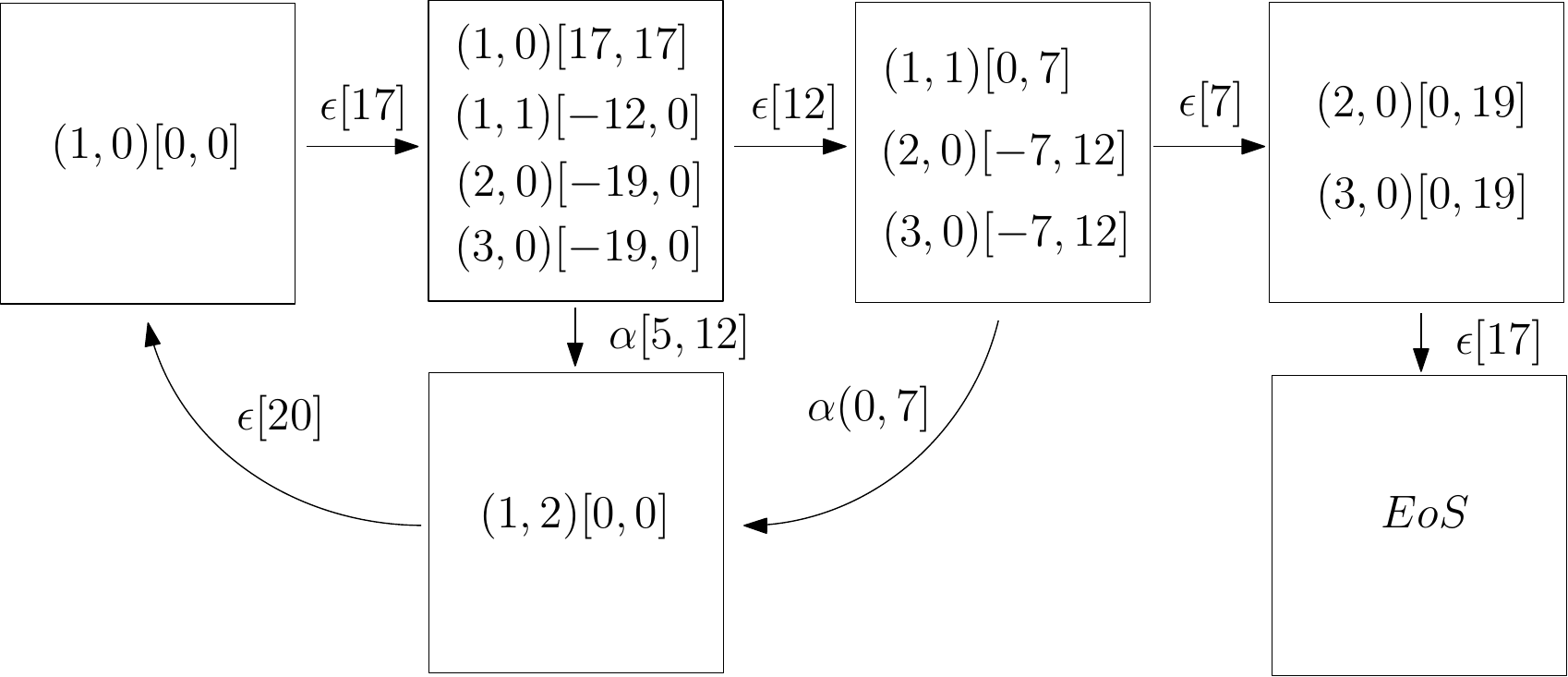}
	\caption{Observer constructed for the timed abstraction in Fig. \ref{fig_ta}.}
	\label{fig_obs}
\end{figure}

The basic observer of the example is shown in Fig. \ref{fig_obs}. The basic observer is constructed as a deterministic finite automaton driven by an external timer and output symbols observed from $H$.

In summary, given a trajectory
simulated from an initial state, the possible discrete state (current location) can be estimated at any time by an observer for any trajectory initiated from a neighborhood around the simulated initial state. There are two notions of time, one is the \textbf{external time} that can be read from an external timer and is reset to zero every time the constructed observer updates its states, the other one is the \textbf{clock time} that is associated with each trajectory which is reset to zero every time the trajectory enters a new location. \textbf{In this paper, we use $t$ to denote the external time and $\tau$ to denote the clock time.} As different trajectories may reach the guards or leave an invariant set at different times, the clock time that is associated with each trajectory has temporal uncertainties. As the clock time is reset to zero when the trajectory enters a new location, the clock time is also associated with each location the trajectory enters. It can be seen that $\tau$ in $\xi_{\ell^m_k}(\tau,x^0_{\ell^m_k})$ is the clock time associated with location $\ell^m_k$ (the location corresponding to the $m$th segment of the $k$th trajectory $\rho_k$). We denote $s$ as the set of possible observer states at the current time. At the external time $t$, we denote $(k,m)[\bar{a},\bar{b}]\in s$ if $\ell^m_k$ is possible as the current location and the clock time $\tau$ in location $\ell^m_k$ has temporal uncertainty $\tau\in[t+\bar{a},t+\bar{b}]$. For example, if the observer state $s^1=\{(1,0)[17,17],(1,1)[-12,0],(2,0)[-19,0],(3,0)[-19,0])\}$, $s^2=\{(1,1)[0,7],(2,0)[-7,12]),(3,0)[-7,12]\}$, and the observer state update is $s^1\xrightarrow[]{\epsilon[12]}s^2$~(here $\epsilon[12]$ means no event is observed for 12 time units), then at external time $t\in[0,12)$, the state could be in location $\ell^1_0$, $\ell^1_1$, $\ell^0_2$ or $\ell^0_3$, the clock time $\tau^1_0$ for location $\ell^1_0$ has no temporal uncertainty $\tau^1_0=t+17$, the clock time $\tau^1_1$ for location $\ell^1_1$ has temporal uncertainty $\tau^1_1\in[t-12,t]$, etc. The external time $t$ is reset to $0$ when $s^1$ is updated by $s^2$ and at the new external time $t$, the state could be in location $\ell^1_1$, $\ell^0_2$ or $\ell^0_3$, the clock time $\tau^1_1$ for location $\ell^1_1$ has temporal uncertainty $\tau^1_2\in[t,t+7]$, the clock time $\tau^0_2$ for location $\ell^0_2$ has temporal uncertainty $\tau^0_2\in[t-7,t+12]$, etc. \textbf{To summarize, after the basic observer is constructed, the external time $t$ has no temporal uncertainties, while the clock time $\tau$ has temporal uncertainties}.

	Note that when $\bar{a}$ in $(k,m)[\bar{a},\bar{b}]$ is negative, it actually represents ``latent'' states that are currently in other locations. For example, $(2,0)[-7,12]$ means at the external time $t$ ($t<7$), the hybrid system state may have already been at location $\ell^0_2$ for $\tau$ time units ($\tau$ is the positive clock time in location $\ell^0_2$, $\tau\in[0,t]$), but may also
	be at location $\ell^1_1$ and will enter location $\ell^0_2$ at the next ($-\tau$) time unit ($\tau$ is the ``virtual'' negative clock time in location $\ell^0_2$, $\tau\in[t-7,0]$). To account for the negative times, we allow $\tau$ in the notation $\xi_{\ell^m}(\tau,x^0_{\ell^m})$ to be negative to represent the ``virtual'' negative clock time when the state is not in location $\ell^m$ at the current time but will enter location $\ell^m$ at a future time $(-\tau)$.
	
	\begin{definition}                                                            
		The time-robust tube segment at external time $t$ corresponding to location $\ell^m_k$ and an interval $[t+\bar{a},t+\bar{b}]$, denoted as $R_{tube}(k,m,[t+\bar{a},t+\bar{b}])$, is defined as follows:
		\begin{align}\nonumber
		\begin{split} 
		& R_{tube}(k,m,[t+\bar{a},t+\bar{b}])=\{\big(\tau,\hat{\xi}_{\ell^m_k}(\tau,\tilde{x}^0_{\ell^m_k})\big)~\vert~\tau\in[t+\bar{a},\\
		&t+\bar{b}],~\xi_{\ell^m_k}(\tau,\tilde{x}^0_{\ell^m_k})\in B_{\ell^m_k}(\gamma_{\ell^m_k},\xi_{\ell^m_k}(\tau,x^0_{\ell^m_k}))~\textrm{if}~\tau\ge 0\},
		\end{split}
		\end{align}    
		\label{tube_def}
	\end{definition}
	As $B_{\ell^m_k}(\gamma_{\ell^m_k},\xi_{\ell^m_k}(\tau,x^0_{\ell^m_k}))$ is obtained through the robust neighborhood approach, the time-robust tube segment can be also expressed as 
	\begin{eqnarray}\nonumber
	& & R_{tube}(k,m,[t+\bar{a},t+\bar{b}]):=\{\big(\tau,\hat{\xi}_{\ell^m_k}(\tau,\tilde{x}^0_{\ell^m_k})\big)~\vert~\\\nonumber& &\tau\in[t+\bar{a},t+\bar{b}],\tilde{x}^0_{\ell^m_k}\in B_{\ell^m_k}(\gamma_{\ell^m_k},x^0_{\ell^m_k})\},
	\label{rtube}
	\end{eqnarray}

\begin{proposition}
	Let $H$ be a hybrid automaton, and $O$ be the constructed basic observer. Given that the current state of $O$ is $s$, and the external time is $t$, then the clock time and the state of $\mathcal{H}$ should be in $\{(\tau_k^m,\ell_k^m,x)\vert (\tau_k^m,x)\in R_{tube}(k,m,[t+\bar{a}, t+\bar{b}]), (k,m)[\bar{a},\bar{b}]\in s\}$.
	
	\begin{proof}
		Directly follow from construction of $O$.
	\end{proof}
\end{proposition}

	\section{Robust Temporal Logic Inference for Classification with Spatial and Temporal Uncertainties}
	\label{sec_stl}
	In this section, we present the robust temporal logic inference framework for classification that accounts for both spatial and temporal uncertainties. We first review the metric temporal logic (MTL) \cite{Donze2010}. 
	
	The continuous state of the system we are studying is described by a set of $n$ variables that can be
	written as a vector $x = \{x_1, x_2, \dots, x_n\}$.
	The domain of $x$ is denoted by $\mathcal{X}$. A set $AP=\{\mu_1,\mu
	_2,\dots \mu_q\}$ is a set of atomic propositions, each mapping $\mathcal{X}$ to $\mathbb{B}$. The
	syntax of MTL is defined recursively as follows:
	\[
	\phi:=\top\mid \mu \mid\neg\phi\mid\phi_{1}\land\phi_{2}\mid\phi_{1}\lor
	\phi_{2}\mid\phi_{1}\mathcal{U}_{I}\phi_{2}%
	\]
	where $\top$ stands for the Boolean constant True, 
	$\neg$ (negation), $\land$(conjunction), $\lor$ (disjunction)
	are standard Boolean connectives, $\mathcal{U}$ is a temporal operator
	representing "until", and $I$ is an interval of the form $I=(i_{1},i_{2}),(i_{1},i_{2}],[i_{1},i_{2})$ or $[i_{1},i_{2}]$, $i_1, i_2\ge 0$. We
	can also derive two useful temporal operators from
	"until" ($\mathcal{U}$), which are "eventually" $\Diamond_I\phi=\top\mathcal{U}_I\phi$ and
	"always" $\Box_I\phi=\neg\Diamond_I\neg\phi$.
	
	For a set $S\subseteq \mathcal{X}$, we define the signed distance from $x$ to $S$ as
	\begin{equation}
	\textbf{Dist$_d(x,S)\triangleq$}
	\begin{cases}
	-\textrm{inf}\{d(x, y)\vert y\in cl(S)\},& \mbox{if $x$ $\not\in S$},\\  
	\textrm{inf}\{d(x, y)\vert y\in \mathcal{\mathcal{X}}\setminus S\}, & \mbox{if $x$ $\in S$}.
	\end{cases}                        
	\label{sign}
	\end{equation}
	where $d$ is a metric on $\mathcal{X}$ and $cl(S)$ denotes the closure of the set $S$. In this paper, we use the metric $d(x,y)=\norm{x-y}$, where $\left\Vert\cdot\right\Vert $ denotes the 2-norm. 
	
	The robustness degree of a MTL formula $\phi$ with respect to a trajectory $\xi_\ell(\tau,x^{0}_\ell)$ 
	at time $\tau$ is denoted as $r(\xi_\ell(\tau,x^{0}_\ell),\phi)$:
	\[
	\begin{split}
	r(\xi_\ell(\tau,x^{0}_\ell),\top):=&\infty,\\
	r(\xi_\ell(\tau,x^{0}_\ell),\mu)
	:=& \textbf{Dist$_d(\xi_\ell(\tau,x^{0}_\ell),\mathcal{O}(\mu))$},\\
	r(\xi_\ell(\tau,x^{0}_\ell),\neg\phi):=&-r(\xi_\ell(\tau,x^{0}_\ell),\phi),\\
	r(\xi_\ell(\tau,x^{0}_\ell),\phi_1\land\phi_2)   
	:=& \min\{r(\xi_\ell(\tau,x^{0}_\ell),\phi_1), r(\xi_\ell(\tau,x^{0}_\ell),\phi_2)\},\\
	r(\xi_\ell(\tau,x^{0}_\ell),\phi_1\mathcal{U}_{I}\phi_2)   
	:=& \max\limits_{\tau'\in \tau+ I} \min\{r(\xi_\ell(\tau',x^{0}_\ell),\phi_2),\\
	&\min\limits_{\tau''\in [\tau,\tau')}r(\xi_\ell(\tau'',x^{0}_\ell),\phi_1)\}.
	\end{split}
	\]

 We denote all the functions (trajectories) mapping from $\mathbb{T}=\mathbb{R}$ to $\mathbb{X}$ as $\mathbb{X}^{\mathbb{T}}$ and denote all the functions (trajectories) mapping from $\mathbb{T}_H$ to $\mathbb{X}$ as $\mathbb{X}^{\mathbb{T}_H}$, where $\mathbb{T}_H\triangleq[0,H]$. 
 
 \begin{definition}
 For a set-valued mapping $\Pi_H:\mathbb{X}^{\mathbb{T}_H}\rightarrow\mathbb{X}^{\mathbb{T}}$, and $\hat{\xi}_\ell(\tau,x^{0}_\ell)$ is defined as the extended trajectory of a trajectory segment $\xi_\ell(\tau,x^{0}_\ell)$, denoted as $\hat{\xi}_\ell(\cdot,x^{0}_\ell)=\Pi_H(\xi_\ell(\cdot,x^{0}_\ell))$, if
 	\begin{equation}
 	\hat{\xi}_\ell(\tau,x^{0}_\ell)=\xi_\ell(\tau,x^{0}_\ell), \forall\tau\ge0.
 	\end{equation} 
 	\label{projection}                               
 \end{definition}                     
 
Next, we introduce the Boolean semantics of an MTL suffix in the strong and the weak view, which are modified from the literature of temporal logic model checking and monitoring \cite{KupfermanVardi2001}. In the following, $\hat{\xi}_{\ell^m}(\tau,x^0_{\ell^m})\models_{S}\phi$ (resp. $\hat{\xi}_{\ell^m}(\tau,x^0_{\ell^m})\models_{W}\phi$)
means the extended trajectory $\hat{\xi}_{\ell^m}(\tau,x^0_{\ell^m})$ strongly (resp. weakly) satisfies $\phi$ at time $t$, $\hat{\xi}_{\ell^m}(\tau,x^0_{\ell^m})\not\models_{S}\phi$ (resp. $\hat{\xi}_{\ell^m}(\tau,x^0_{\ell^m})\not\models_{W}\phi$)
means the extended trajectory $\hat{\xi}_{\ell^m}(\tau,x^0_{\ell^m})$ fails to strongly (resp. weakly) satisfy $\phi$ at time $t$.

\begin{definition}
	The Boolean semantics of the (F,G)-fragment MTL for the extended trajectories in the strong view is defined recursively as follows:
	\[
	\begin{split}
	\hat{\xi}_{\ell^m}(\tau,x^0_{\ell^m})\models_{S}\mu\quad\mbox{iff}\quad& \tau\ge 0~\mbox{and}~\\&\textbf{Dist$_d(\xi_{\ell^m}(\tau,x^0_{\ell^m}),\mathcal{O}(\mu))>0$},\\
	\hat{\xi}_{\ell^m}(\tau,x^0_{\ell^m})\models_{S}\lnot\phi\quad\mbox{iff}\quad & \hat{\xi}_{\ell^m}(\tau,x^0_{\ell^m})\not\models_{W}\phi,\\
	\hat{\xi}_{\ell^m}(\tau,x^0_{\ell^m})\models_{S}\phi_{1}\wedge\phi_{2}\quad\mbox{iff}\quad &  \hat{\xi}_{\ell^m}(\tau,x^0_{\ell^m})\models_{S}\phi
	_{1}~\mbox{and}~\\&\hat{\xi}_{\ell^m}(\tau,x^0_{\ell^m})\models_{S}\phi_{2},\\
	\hat{\xi}_{\ell^m}(\tau,x^0_{\ell^m})\models_{S}\phi_{1}\vee\phi_{2}\quad\mbox{iff}\quad &  \hat{\xi}_{\ell^m}(\tau,x^0_{\ell^m})\models_{S}\phi
	_{1}~\mbox{or}~\\&\hat{\xi}_{\ell^m}(\tau,x^0_{\ell^m})\models_{S}\phi_{2},\\	
	\hat{\xi}_{\ell^m}(\tau,x^0_{\ell^m})\models_{S}F_{[\tau_1,\tau_2)}\phi\quad\mbox{iff}\quad &  \exists
	\tau^{\prime}\in[\tau+\tau_1,\tau+\tau_2),\\
	& s.\tau.~\hat{\xi}_{\ell^m}(\tau',x^{0}_{\ell^m})\models_{S}\phi,\\
	\hat{\xi}_{\ell^m}(\tau,x^0_{\ell^m})\models_{S}G_{[\tau_1,\tau_2)}\phi\quad\mbox{iff}\quad &  \hat{\xi}_{\ell^m}(\tau',x^{0}_{\ell^m})\models_{S}\phi, \\&\forall
	\tau^{\prime}\in[\tau+\tau_1, \tau+\tau_2). 
	\end{split}
	\]
	\label{strong}
\end{definition}

\begin{definition}
	The Boolean semantics of the (F,G)-fragment MTL for the extended trajectories in the weak view is defined recursively as follows:
	\[
	\begin{split}
	\hat{\xi}_{\ell^m}(\tau,x^0_{\ell^m})\models_{W}\mu\quad\mbox{iff}\quad& \textrm{either of the following holds}:\\
	& 1)~\tau\ge0~\mbox{and}~\\
	&\textbf{Dist$_d(\xi_{\ell^m}(\tau,x^0_{\ell^m}),\mathcal{O}(\mu))>0$};\\
	& 2)~\tau<0,\\
	\hat{\xi}_{\ell^m}(\tau,x^0_{\ell^m})\models_{W}\lnot\phi\quad\mbox{iff}\quad & \hat{\xi}_{\ell^m}(\tau,x^0_{\ell^m})\not\models_{S}\phi,\\
	\hat{\xi}_{\ell^m}(\tau,x^0_{\ell^m})\models_{W}\phi_{1}\wedge\phi_{2}\quad\mbox{iff}\quad &  \hat{\xi}_{\ell^m}(\tau,x^0_{\ell^m})\models_{W}\phi
	_{1}~\mbox{and}\\&~\hat{\xi}_{\ell^m}(\tau,x^0_{\ell^m})\models_{W}\phi_{2},\\
	\hat{\xi}_{\ell^m}(\tau,x^0_{\ell^m})\models_{W}\phi_{1}\vee\phi_{2}\quad\mbox{iff}\quad &  \hat{\xi}_{\ell^m}(\tau,x^0_{\ell^m})\models_{W}\phi
	_{1}~\mbox{or}\\&~\hat{\xi}_{\ell^m}(\tau,x^0_{\ell^m})\models_{W}\phi_{2},\\
	\hat{\xi}_{\ell^m}(\tau,x^0_{\ell^m})\models_{W}F_{[\tau_1,\tau_2)}\phi\quad\mbox{iff}\quad & \exists
	\tau^{\prime}\in[\tau+\tau_1,\tau+\tau_2), \\
	& s.t.~\hat{\xi}_{\ell^m}(\tau',x^{0}_{\ell^m})\models_{W}\phi,\\
	\hat{\xi}_{\ell^m}(\tau,x^0_{\ell^m})\models_{W}G_{[\tau_1,\tau_2)}\phi\quad\mbox{iff}\quad & \hat{\xi}_{\ell^m}(\tau',x^{0}_{\ell^m})\models_{W}\phi,\\& \forall
	\tau^{\prime}\in[\tau+\tau_1, \tau+\tau_2).
	\label{weak}
	\end{split}
	\]
\end{definition}

\begin{figure}[th]
	\centering
	\includegraphics[width=8cm]{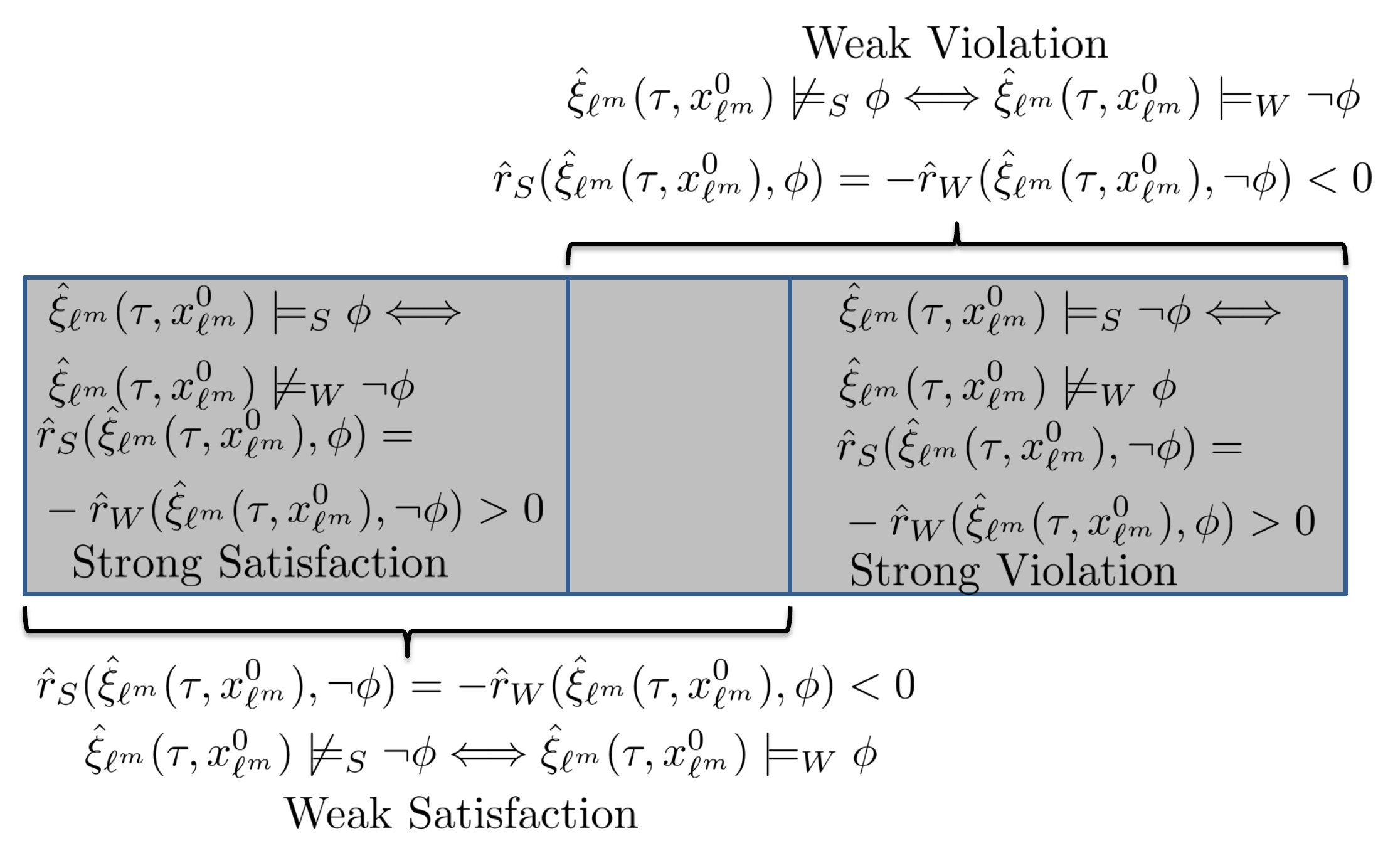}\caption{Venn diagram of strong (weak) satisfaction and strong (weak) violation.}
	\label{view}
\end{figure}
	
	We denote $\hat{r}_S(\hat{\xi}_{\ell^m}(\tau,x^0_{\ell^m}),\phi)$ and $\hat{r}_W(\hat{\xi}_{\ell^m}(\tau,x^0_{\ell^m}),\phi)$ as the extended robustness degree of an extended trajectory $\hat{\xi}_{\ell^m}(\tau,x^0_{\ell^m})$ with respect to an MTL formula $\phi$ evaluated at a certain external time corresponding to the clock time $\tau$ for the $i$th trajectory ($\tau$ can be positive or negative) in the strong and the weak view, respectively. $\hat{r}_S(\hat{\xi}_{\ell^m}(\tau,x^0_{\ell^m}),\phi)$ can be calculated recursively via the following extended quantitative semantics:             
   \begin{align}  
   \begin{split}
   & \hat{r}_S(\hat{\xi}_{\ell^m}(\tau,x^0_{\ell^m}),\mu)  =	
   \begin{cases}
   r(\xi_{\ell^m}(\tau,x^0_{\ell^m}),\mu),~~\mbox{if $\tau\ge0$},\\    
   -\infty, ~~\mbox{if $\tau<0$},
   \end{cases}  \\
   &\hat{r}_S(\hat{\xi}_{\ell^m}(\tau,x^0_{\ell^m}),\lnot\phi)    =-\hat{r}_W(\hat{\xi}_{\ell^m}(\tau,x^0_{\ell^m}),\phi),\\
   &\hat{r}_S(\hat{\xi}_{\ell^m}(\tau,x^0_{\ell^m}),\phi_{1}\wedge\phi_{2})  =  \min(\hat{r}_S(\hat{\xi}_{\ell^m}(\tau,x^0_{\ell^m}),\phi_{1}),\\&~~~~~~~~~~~~~~~~~~~~~~~~~~~~~~~~~ \hat{r}_S(\hat{\xi}_{\ell^m}(\tau,x^0_{\ell^m}),\phi_{2})),\\	
   &\hat{r}_S(\hat{\xi}_{\ell^m}(\tau,x^{0}_{\ell^m}),F_{I}\phi)   
   := \max\limits_{\tau'\in (\tau+I)} \hat{r}_S(\hat{\xi}_{\ell^m}(\tau',x^{0}_{\ell^m}),\phi),\\
   &\hat{r}_S(\hat{\xi}_{\ell^m}(\tau,x^{0}_{\ell^m}),G_{I}\phi)   
   := \min\limits_{\tau'\in (\tau+I)} \hat{r}_S(\hat{\xi}_{\ell^m}(\tau',x^{0}_{\ell^m}),\phi).
   \end{split}
   \label{semantics}
   \end{align}  
  $\hat{r}_W(\hat{\xi}_{\ell^m}(\tau,x^0_{\ell^m}),\phi)$ can be calculated recursively via the following extended quantitative semantics:     
  \begin{align}  
  \begin{split}
  & \hat{r}_W(\hat{\xi}_{\ell^m}(\tau,x^0_{\ell^m}),\mu)  =	
  \begin{cases}
  r(\xi_{\ell^m}(\tau,x^0_{\ell^m}),\mu),~~\mbox{if $\tau\ge0$},\\    
  \infty, ~~\mbox{if $\tau<0$},
  \end{cases}  \\
  &\hat{r}_W(\hat{\xi}_{\ell^m}(\tau,x^0_{\ell^m}),\lnot\phi)    =-\hat{r}_S(\hat{\xi}_{\ell^m}(\tau,x^0_{\ell^m}),\phi),\\
  &\hat{r}_W(\hat{\xi}_{\ell^m}(\tau,x^0_{\ell^m}),\phi_{1}\wedge\phi_{2})  =  \min(\hat{r}_W(\hat{\xi}_{\ell^m}(\tau,x^0_{\ell^m}),\phi_{1}),\\&~~~~~~~~~~~~~~~~~~~~~~~~~~~~~~~~~ \hat{r}_W(\hat{\xi}_{\ell^m}(\tau,x^0_{\ell^m}),\phi_{2})),\\	
  &\hat{r}_W(\hat{\xi}_{\ell^m}(\tau,x^{0}_{\ell^m}),F_{I}\phi)   
  := \max\limits_{\tau'\in (\tau+I)} \hat{r}_W(\hat{\xi}_{\ell^m}(\tau',x^{0}_{\ell^m}),\phi),\\
  &\hat{r}_W(\hat{\xi}_{\ell^m}(\tau,x^{0}_{\ell^m}),G_{I}\phi)   
  := \min\limits_{\tau'\in (\tau+I)} \hat{r}_W(\hat{\xi}_{\ell^m}(\tau',x^{0}_{\ell^m}),\phi).
  \end{split}
  \label{semantics2}
  \end{align}

	\begin{definition}     
		Given a labeled set of extended trajectories $\{(\hat{\xi}_{\ell^{m_i}_{k_i}}(\tau_i,x^{0}_{\ell^{m_i}_{k_i}}),c_i)\}^{N}_{i=1}$ from a hybrid system $\mathcal{H}$, $c_i=1$ represents desired behavior and $c_i=-1$ represents undesired behavior, an MTL formula $\phi$ evaluated at external time $t$ (corresponding to clock time $\tau_i$ for the $i$th trajectory, each $\tau_i$ can be positive or negative) perfectly classifies the desired behaviors (trajectory segments) and undesired behaviors (trajectory segments) if the following condition is satisfied:\\
		$\hat{\xi}_{\ell^{m_i}_{k_i}}(\tau_i,x^{0}_{\ell^{m_i}_{k_i}})\models_{W}\phi$, if $c_i=1$; $\hat{\xi}_{\ell^{m_i}_{k_i}}(\tau_i,x^{0}_{\ell^{m_i}_{k_i}})\models_{W}\lnot\phi$, if $c_i=-1$.
		\label{perfect0}
	\end{definition}

	\begin{problem}
		Given a labeled set of time-robust tube segments $\tilde{S}=\{(R_{tube}(k_i,m_i,[t+\bar{a}_i,t+\bar{b}_i]),c_i)\}^{N}_{i=1}$ from a hybrid system $\mathcal{H}$, find an MTL formula $\phi$ such that $\phi$ evaluated at external time $t$ (corresponding to different clock times for different trajectory segments) perfectly classifies the desired behaviors (trajectory segments) and undesired behaviors (trajectory segments) in $\tilde{S}$, i.e. for any $\big(\tau_i,\hat{\xi}_{\ell^{m_i}_{k_i}}(\tau_i,\tilde{x}^0_{\ell^{m_i}_{k_i}})\big)\in R_{tube}(k_i,m_i,[t+\bar{a}_i,t+\bar{b}_i])$, $\hat{r}_W(\hat{\xi}_{\ell^{m_i}_{k_i}}(\tau_i,\tilde{x}^0_{\ell^{m_i}_{k_i}}),\phi)>0$, if $c_i=1$;  $\hat{r}_W(\hat{\xi}_{\ell^{m_i}_{k_i}}(\tau_i,\tilde{x}^0_{\ell^{m_i}_{k_i}}),\lnot\phi)>0$, if $c_i=-1$.
		\label{Problem}
	\end{problem}

	If for each location $\ell$, the continuous dynamics is affine and stable, then there exists a quadratic autobisimulation function $\Phi_\ell(\xi_\ell(\tau,x^{0}_\ell), \xi_\ell(\tau,x)) =  [\big(\xi_\ell(\tau,x^0_\ell)-\xi_\ell(\tau,x)\big)^TM_\ell\big(\xi_\ell(\tau,x^{0}_\ell)-\xi_\ell(\tau,x)\big)]^{\frac{1}{2}}$, where $M_\ell$ is positive definite. To solve problem \ref{Problem}, we first give the following three propositions:	
	\begin{proposition}
		\label{prop_space}
		For any MTL formula $\phi$ and $\gamma_{\ell}>0$, if $\Phi_\ell(\xi_{\ell}(\tau,\tilde{x}^0_\ell),\xi_{\ell}(\tau,x^0_\ell))=[\big(\xi_\ell(\tau,x^0_\ell)-\xi_\ell(\tau,\tilde{x}^0_\ell)\big)^TM_\ell$ $\big(\xi_\ell(\tau,x^{0}_\ell)-\xi_\ell(\tau,\tilde{x}^0_\ell)\big)]^{\frac{1}{2}}<\gamma_{\ell}$ for any $\tau\ge0$, then for any $\tau$ (here $\tau$ can be positive or negative), we have
		\begin{align} 
			\begin{split} 
		&\hat{r}_S(\hat{\xi}_{\ell}(\tau,x^0_\ell),\phi)-\hat{\gamma}_{\ell}\le \hat{r}_S(\hat{\xi}_{\ell}(\tau,\tilde{x}^0_\ell),\phi)\\&\le \hat{r}_S(\hat{\xi}_{\ell}(\tau,x^0_\ell),\phi)+\hat{\gamma}_{\ell},\\
		&\hat{r}_W(\hat{\xi}_{\ell}(\tau,x^0_\ell),\phi)-\hat{\gamma}_{\ell}\le \hat{r}_W(\hat{\xi}_{\ell}(\tau,\tilde{x}^0_\ell),\phi)\\&\le \hat{r}_W(\hat{\xi}_{\ell}(\tau,x^0_\ell),\phi)+\hat{\gamma}_{\ell},
	    	\end{split} 
		\end{align}  
		where $c$ is a classification label, $\hat{\gamma}_{\ell}=\gamma_{\ell}\norm{M_\ell^{-\frac{1}{2}}}$ .
	\end{proposition}
	
	\begin{proof} 
		See Appendix.
	\end{proof}
	
	\begin{proposition}	 
		For any MTL formula $\phi$ that only contains one variable $x_j$ ($j=1,2,\dots,n$) and $\gamma_{\ell}$ $>0$, if  $\Phi_\ell(\xi_{\ell}(\tau,\tilde{x}^0_\ell),\xi_{\ell}(\tau,x^0_\ell))=[\big(\xi_\ell(\tau,x^0_\ell)-\xi_\ell(\tau,\tilde{x}^0_\ell)\big)^TM_\ell$ $\big(\xi_\ell(\tau,x^{0}_\ell)-\xi_\ell(\tau,\tilde{x}^0_\ell)\big)]^{\frac{1}{2}}<\gamma_{\ell}$ for any $\tau\ge0$, and if there exists $z_{\ell,j}>0$ such that $z_{\ell,j}^2e_{j}^{T}e_{j} \preceq M_{\ell}$, then for any $\tau$ (here $\tau$ can be positive or negative), we have
		\begin{align} 
		\begin{split} 
		&\hat{r}_S(\hat{\xi}_{\ell}(\tau,x^0_\ell),\phi)-\tilde{\gamma}_{\ell,j}\le \hat{r}_S(\hat{\xi}_{\ell}(\tau,\tilde{x}^0_\ell),\phi)\le\\
		& \hat{r}_S(\hat{\xi}_{\ell}(\tau,x^0_\ell),\phi)+\tilde{\gamma}_{\ell,j},\\
		&\hat{r}_W(\hat{\xi}_{\ell}(\tau,x^0_\ell),\phi)-\tilde{\gamma}_{\ell,j}\le \hat{r}_W(\hat{\xi}_{\ell}(\tau,\tilde{x}^0_\ell),\phi)\le\\
		& \hat{r}_W(\hat{\xi}_{\ell}(\tau,x^0_\ell),\phi)+\tilde{\gamma}_{\ell,j},
		\end{split} 
		\end{align}  
       where $c$ is the classification label, $\tilde{\gamma}_{\ell,j}=\gamma_{\ell}/z_{\ell,j}$.                                                            
		\label{th1}
	\end{proposition}
	
	\begin{proof} 
		See Appendix.
	\end{proof} 
	\begin{remark}	 
		Proposition \ref{th1} provides a possibly tighter bound $\tilde{\gamma}_{\ell,j}$ than $\hat{\gamma}_{\ell}$ when the MTL formula $\phi$ only contains one variable $x_j$, which applies to the case of smart building occupancy detection in Section \ref{sec_implementation} where fewer numbers of applied sensors (corresponding to the number of variables that are contained in $\phi$) are preferred.
	\end{remark}		
	
	\begin{proposition}  
		Given the settings of Problem \ref{Problem}, an MTL formula $\phi$ evaluated at external time $t$ perfectly classifies the desired behaviors and undesired behaviors in $\tilde{S}$ if the following condition is satisfied:\\ $MG(k_i,m_i,\bar{a}_i,\bar{b}_i,\phi,c_i)>0$, if $c_i=1$; $MG(k_i,m_i,\bar{a}_i,\bar{b}_i,$ $\phi,c_i)<0$, if $c_i=-1$, where $MG(\cdot)$ is a margin function defined as follows:
		\begin{align}
		\begin{split}
		MG(k,m,\bar{a},\bar{b},\phi,1)&=\min\limits_{\tau\in t+ [\bar{a},\bar{b}]}\hat{r}_W(\hat{\xi}_{\ell_k^m}(\tau,x^0_{\ell^{m}_{k}}),\phi)-\hat{\gamma}_{\ell^m_k},\\
		MG(k,m,\bar{a},\bar{b},\phi,-1)&=\min\limits_{\tau\in t+ [\bar{a},\bar{b}]}\hat{r}_W(\hat{\xi}_{\ell_k^m}(\tau,x^0_{\ell^{m}_{k}}),\lnot\phi)-\hat{\gamma}_{\ell^m_k},
		\end{split}
		\label{MG}
		\end{align} 
		where $\hat{\gamma}_{\ell_k^m}=\gamma_{\ell_k^m}\norm{M_{\ell_k^m}^{-\frac{1}{2}}}$.
		\label{sol_th}
	\end{proposition}	
	\begin{proof} 
		See Appendix.
	\end{proof}

	According to Proposition \ref{sol_th}, we can solve Problem \ref{Problem} by minimizing the following cost function:
	\begin{align}
	\begin{split}
	\label{eq_cost}
	J(\tilde{S}, \phi)=& \sum_{c_i=1}G(k_i,m_i,\bar{a}_i,\bar{b}_i,\phi,c_i)+\\& \sum_{c_i=-1}G(k_i,m_i,\bar{a}_i,\bar{b}_i,\neg\phi,c_i),
	\end{split}
	\end{align}
	where $G(\cdot)$ is defined as follows:
	\begin{eqnarray}\nonumber
	G(k,m,\bar{a},\bar{b},\phi,c)&=&
	\begin{cases}
	0, \text{ if } MG(k,m,\bar{a},\bar{b},\phi,c)>0,\\
	\zeta, \text{ otherwise},
	\end{cases}
	\label{eq_cost1}
	\end{eqnarray}
	where the margin function $MG(\cdot)$ is defined in (\ref{MG}), $\zeta$ is a positive constant, the external time $t$ is usually set to be $0$. When the MTL formula $\phi$ only contains one variable $x_j$, $\hat{\gamma}_{\ell}$ can be replaced by $\tilde{\gamma}_{\ell,j}$ in (\ref{eq_cost1}).
	
	The core of the classification process is a non-convex optimization problem for finding the structure and the parameters that describe the MTL formula $\phi$, which can be solved through Particle Swarm Optimization~\cite{Kennedy1995}. The search starts from a basis of candidate formulae in the form of $\Box_{\lbrack \tau_{1},\tau_{2}]}\pi$ or $\Diamond_{\lbrack \tau_{1},\tau_{2}]}\pi$ and adding Boolean connectives until a satisfactory formula is found.
	
	Once the optimization procedure obtains an optimal formula $\phi$, we can use the obtained $\phi$ to refine the basic observer constructed as in \cite{YiCDC}. The refinement procedure is to shrink the observer's state (i.e., the state estimate for $\mathcal{H}$) and the subsequent transitions based on satisfaction or violation of a MTL formula. For a given $\phi$, we can shrink the observer state as soon as $\phi$ is satisfied or violated, while not resetting the timer. Then the subsequent states and transitions can be modeled in the same way as the basic observer as constructed in \cite{YiCDC}. 
	
	\section{Implementation}
	\label{sec_implementation} 
\begin{figure}[h]
		\centering
		\includegraphics[scale=0.25]{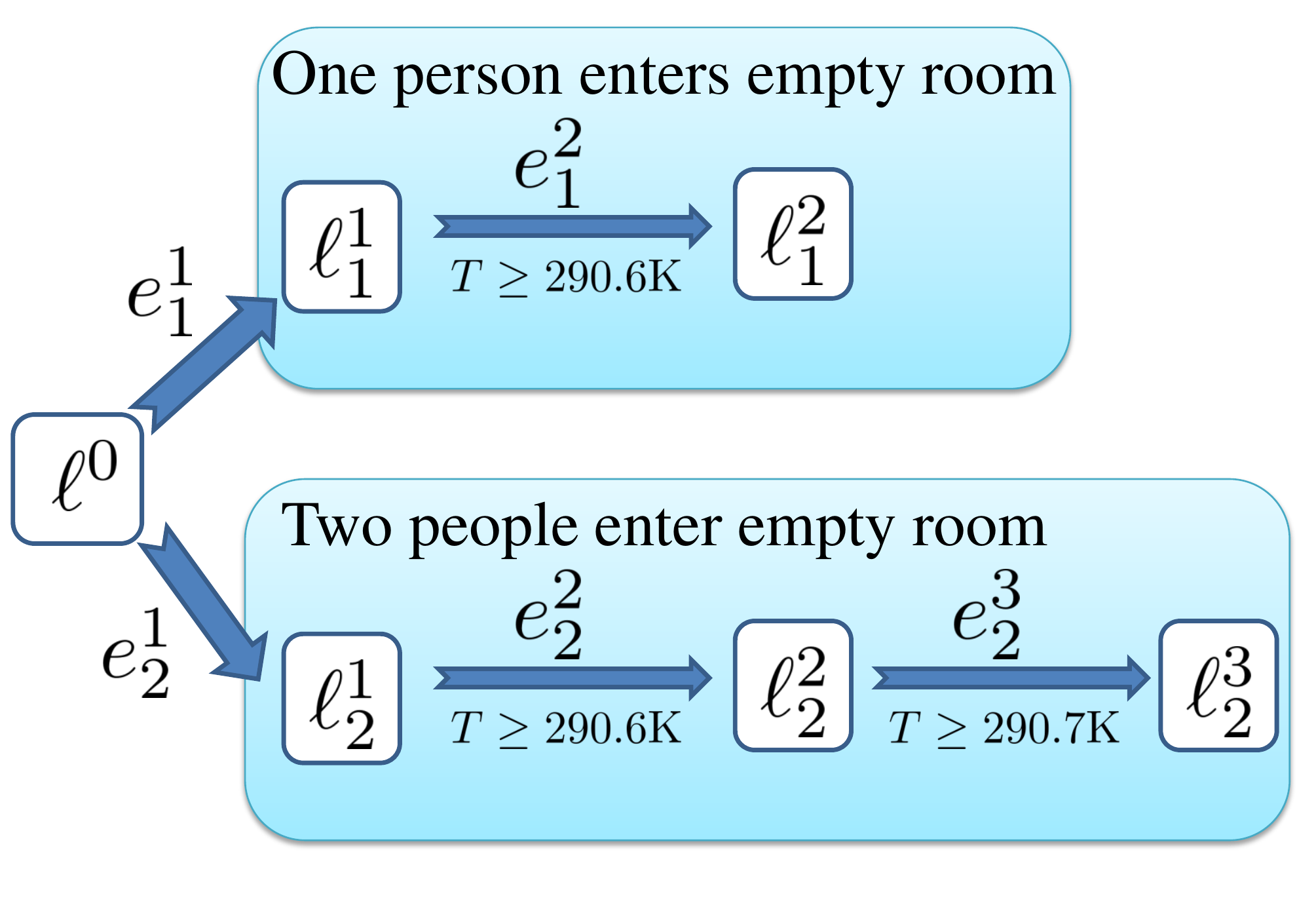}
		\caption{Locations of hybrid system $\mathcal{H}$ for the smart building model describing the series of events of the two cases.} \label{automaton}
	\end{figure}

	In this section, we implement our occupancy detection method to distinguish between two cases in the simulation model of a smart building testbed \cite{Okaeme2016}: (i) one person enters an empty room after the door opens; (ii) two people enter an empty room after the door opens. We assume that we can observe the event when the door opens. The air conditioning is programmed to increase the mass flow rate of the cooling air when the temperature reaches certain thresholds (e.g. 290.6K, 290.7K). 
	The system is modeled as a hybrid system $\mathcal{H}$ with 6 locations, as shown in Fig. \ref{automaton}. The state $x=[T, w, \dot{W}_{\rm{gen}}, \dot{Q}_{\rm{gen}}]$ represents the temperature and humidity ratio of the room, humidity and heat generation rate within the room (i.e. from the humans) respectively (we choose the units of $\dot{W}_{\rm{gen}}$ and $\dot{Q}_{\rm{gen}}$ to be W and mg/s, respectively). $\dot{W}_{\rm{gen}}$ and $\dot{Q}_{\rm{gen}}$ are added as two pseudo-states to account for the variations of the humidity and heat generation rates by different people \cite{TenWolde2007}. The continuous dynamics in the 6 locations are given as follows:\\
	
	For location $\ell^0$ (room unoccupied):                                                              
	\[
	\begin{cases}
	C\dot{x}_1=\dot{m}_{\ell^0}C_{\rm{p}}(T_{\rm{s}}-x_1)+\beta G(x_2-w_{\infty})-K(x_1-T_{\infty});                    \\
	M\dot{x}_2=\dot{m}(w_{\rm{s}}-x_2)-G(x_2-w_{\infty});\\
	\dot{x}_3=0;
	\dot{x}_4=0.
	\end{cases}                 
	\]
	
	For the other 5 locations $\ell_k^m$ ($\ell^1_1$, $\ell^2_1$, $\ell^1_2$, $\ell^2_2$, $\ell^3_2$, room occupied with one or two people):
	\[%
	\begin{cases}
	C\dot{x}_1=\dot{m}_{\ell_k^m}C_{\rm{p}}(T_{\rm{s}}-x_1)+\beta G(x_2-w_{\infty})-K(x_1-T_{\infty})\\
	~~~~~~~~+x_4-10^{-6}\beta x_3;                    \\
	M\dot{x}_2=\dot{m}_{\ell_k^m}(w_{\rm{s}}-x_2)-G(x_2-w_{\infty})+x_3;\\
	\dot{x}_3=0;
	\dot{x}_4=0.
	\end{cases}                 
	\]	
	where $\dot{m}_{\ell_k^m}$ is the mass flow rate of the air conditioning in location $\ell_k^m$ (we set $\dot{m}_{\ell^0}=\dot{m}_{\ell^1_1}=\dot{m}_{\ell^1_2}=0.5$Kg/s$, \dot{m}_{\ell^2_1}=\dot{m}_{\ell^2_2}=0.6$Kg/s$, \dot{m}_{\ell^3_2}=0.8$Kg/s), $C$ is the thermal capacitance of the room, $M$ is mass of air in the room, $G$ is the mass transfer conductance between the room
	and the ambient, $w_{\rm{s}}$,
	$T_{\rm{s}}$ are the supply air humidity ratio and temperature respectively, $w_{\infty}$,
	$T_{\infty}$ are the ambient humidity ratio and temperature respectively, $C_{\rm{p}}$ is specific heat of air at constant pressure, $\beta$ is latent heat of vaporization of water,
	$K$ is the wall thermal conductance.
	
\begin{figure}
		\centering
		\includegraphics[scale=0.3]{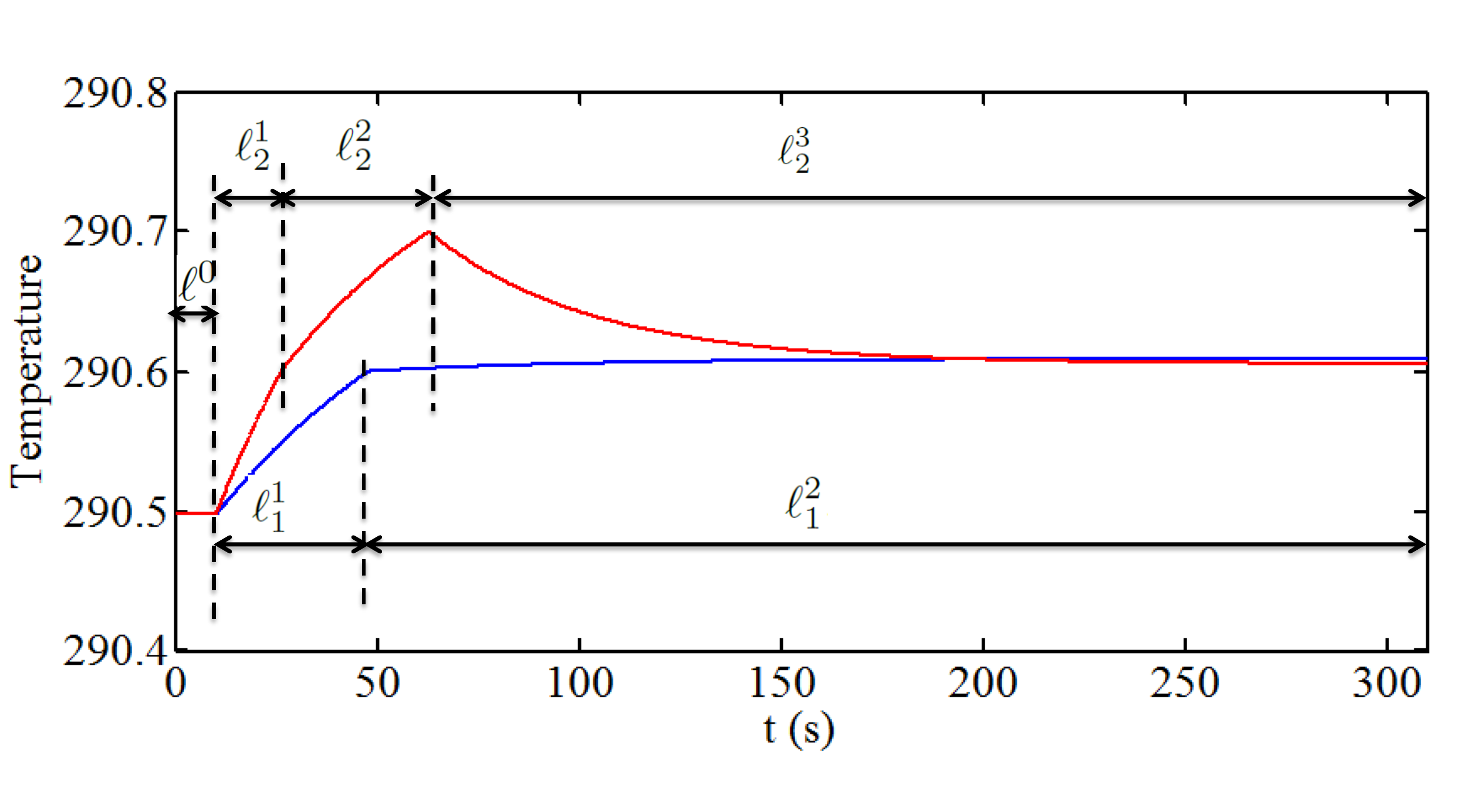}
		\caption{The temperature state of the two simulated trajectories (blue represents the trajectory when one person enters the empty room, red represents the trajectory when two people enter the empty room) and the corresponding locations.} 
		\label{pic}
	\end{figure}

	We set $T_{\infty}=303$K~(29.85$^{\circ}$C), $T_{\rm{s}}=290$K~(16.85$^{\circ}$C), $w_{\infty}=0.0105$, $w_{\rm{s}}=0.01$. As shown in Fig. \ref{pic}, as human can generate both heat and moisture, the room temperature and humidity ratio will increase towards the new equilibrium after people enter the room. As the mass flow rate of the air conditioning may change in different locations, when two people enter the empty room, the temperature first increases to 290.7K, then starts to decrease as the mass flow rate is increased to 0.8Kg/s. It can be seen that the steady state values of the temperatures in the two cases are almost the same, therefore a temporal logic formula is needed to distinguish their temporal patterns in the transient period. 	

	\begin{figure}
		\centering
		\includegraphics[scale=0.5]{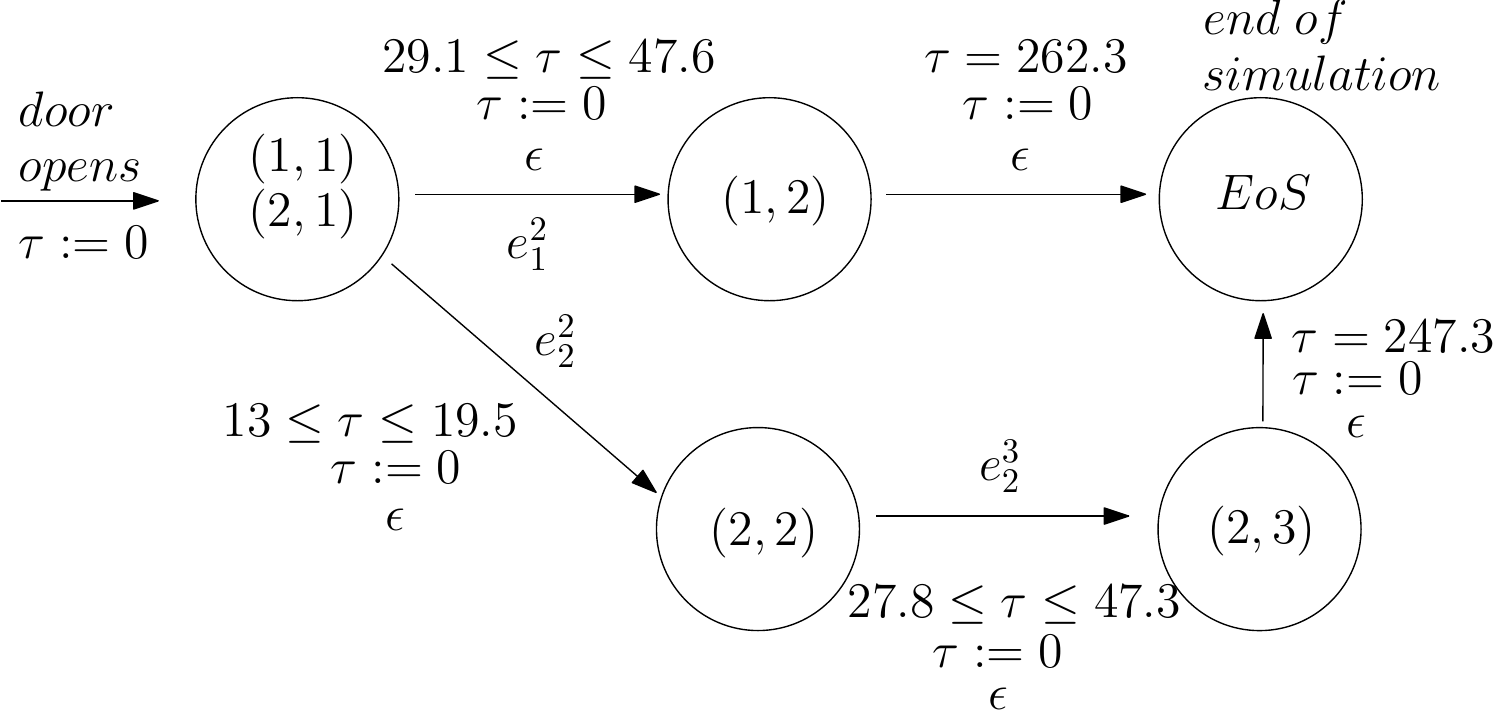}
		\caption{A timed abstraction of the hybrid automaton $\mathcal{H}$. $\tau$ is the clock time that is associated with each trajectory which is reset to zero every time the trajectory enters a new location. For instance, the transition from $(1,1)$ to $(1,2)$ means that any trajectory of $\mathcal{H}$ initiated from $B_{\ell^1_1}(\gamma_{\ell^1_1},x^0_{\ell^1_1})$ will reach $B_{\ell^2_1}(\gamma_{\ell^2_1},x^0_{\ell^2_1})$ within $29.1$ to $47.6$ time units by triggering an unobservable event.} 
		\label{fig_ta_exp}
	\end{figure}
	
	The invariant sets are 
	\begin{eqnarray*}
		Inv(\ell^0) &=&\mathbb{R}^4,\\
		Inv(\ell^1_1) &=&Inv(\ell^1_2)= \{x\vert 290.4\le x_2\le 290.6\},\\
		Inv(\ell^2_1) &=&Inv(\ell^2_2)= \{x\vert 290.5\le x_2\le 290.7\},\\
		& &Inv(\ell^3_2)= \{x\vert 290.6\le x_2\le 290.8\}.
	\end{eqnarray*}
	The events are modeled as follows:
	\begin{itemize}
		\item $e^1_1 =(\ell^0,\ell^1_1,g^1_1,r^1_1)$, where $g^1_1=
		\mathbb{R}^4$, $r^1_1(x)=x+[0,0,80,300]$;	
		\item $e^1_2 =(\ell^0,\ell^1_2,g^1_2,r^1_2)$, where $g^1_2=
		\mathbb{R}^4$, $r^1_2(x)=x+[0,0,160,600]$;     	
		\item $e^2_1=(\ell^1_1,\ell^2_1,g^2_1,r^2_1) = e^2_2=(\ell^1_2,\ell^2_2,g^2_2,r^2_2)$, where $g^2_1=g^2_2=\{x\vert
		x_2=290.6\}$, $r^2_1(x)=r^2_2(x)=x$;
		\item $e^3_2=(\ell^2_2,\ell^3_2,g^3_2,r^3_2)$, where $g^3_2=\{x\vert
		x_2=290.7\}$, $r^3_2(x)=x$.
	\end{itemize}
	
	The events $e^1_1$ and $e^1_2$ are non-deterministic, i.e. the events can happen anywhere in $Inv(\ell^0)$; the events $e^2_1$, $e^2_2$ and $e^3_2$ are deterministic, i.e. the events are forced to occur whenever the states leave the invariant sets (reach the guards). The output symbols of events $e^1_1$ and $e^1_2$ are observable (door opening) while the output symbols of events $e^2_1$, $e^2_2$ and $e^3_2$ are unobservable.
	
	The reset initial state at location $\ell^1_1$ lies in the following set:
	\begin{eqnarray*}
		\mathcal{L}^1_1\times \mathcal{X}^1_1=\{\ell^1_1\}\times \{x&\vert& x_1=0.01, x_2=290.4976, \\
		& & 280\le x_3\le 320, 60\le x_4\le 100\}.
	\end{eqnarray*}  
	The reset initial state at location $\ell^1_2$ lies in the following set:
	\begin{eqnarray*}
		\mathcal{L}^1_2\times \mathcal{X}^1_2=\{\ell^1_1\}\times \{x&\vert& x_1=0.01, x_2=290.4976, \\
		& &560\le x_3\le 640, 120\le x_4\le 200\}.
	\end{eqnarray*}  
	By using the MATLAB Toolbox STRONG~\cite{Deng2013}, we can verify       
	that $\mathcal{L}^1_1\times \mathcal{X}^1_1$ is covered by a robust neighborhood
	$B_{\ell^1_1}(\gamma_{\ell^1_1},x^0_{\ell^1_1})=\{x\vert \Phi_{\ell^1_1}(x,x^0_{\ell^1_1})<\gamma_{\ell^1_1}=0.098\}$
	around the reset initial state $x^0_{\ell^1_1}$ of the trajectory (w.l.o.g, we assume that the door opens at 10 seconds)
	\[
	\begin{split}	
	\rho_1 =& (e^0, \ell^0, x^0_{\ell^0}, 10),(e^1_1, \ell^1_1, x^0_{\ell^1_1}, 37.6), (e^2_1, \ell^2_1, x^0_{\ell^2_1}, 262.4),\\
	x^0_{\ell^0}   =& [0.01, 290.4976,0,0]^T,
	x^0_{\ell^1_1} = [0.01,290.4976,300,80]^T,\\
	x^0_{\ell^2_1} =& [0.0101,290.6,300,80]^T.	
	\end{split}	
	\]
	
	Similarly, we can verify       
	that $\mathcal{L}^1_2\times \mathcal{X}^1_2$ is covered by a robust neighborhood
	$B_{\ell^1_2}(\gamma_{\ell^1_2},x^0_{\ell^1_2})=\{x\vert \Phi_{\ell^1_2}(x,x^0_{\ell^1_2})<\gamma_{\ell^1_2}=0.1\}$
	around the reset initial state $x^0_{\ell^1_2}$ of the trajectory
	\[
	\begin{split}
	\rho_2 =& (e^0, \ell^0, x^0_{\ell^0}, 10),(e^1_2, \ell^1_2, x^0_{\ell^1_2}, 16.1), (e^2_2, \ell^2_2, x^0_{\ell^2_2}, 36.6),\\& (e^3_2, \ell^3_2, x^0_{\ell^3_2}, 247.3),\\
	x^0_{\ell^0}   =& [0.01, 290.4976,0,0]^T,
	x^0_{\ell^1_2} = [0.01, 290.4976, 600, 160]^T,\\
	x^0_{\ell^2_2} =& [0.0101, 290.6, 600, 160]^T,\\
	x^0_{\ell^3_2} =& [0.0102, 290.7, 600, 160]^T.
	\end{split}
	\]	
	
	%	\begin{figure}
	%		\centering
	%		\includegraphics[scale=0.35]{ellipseperson.pdf}
	%		\caption{Outer bounds of states $T, \dot{Q}_{gen}$ in the ellipsoid $B_{\ell}(\gamma,x)$.} \label{ellipse}
	%	\end{figure}
	
	As the variation range of the temperature is much smaller than the variation ranges of the humidity and heat generation rates, in order to cover the reset initial sets $\mathcal{L}^1_1\times\mathcal{X}^1_1$ and $\mathcal{L}^1_2\times\mathcal{X}^1_2$, we optimize the matrix $M_{\ell}$ in each location $\ell$ (geometrically change the shape of the level set ellipsoid) so that the outer bounds of the level set ellipsoid $B_{\ell}(\gamma_\ell,x^0_\ell)$ in the dimension of the temperature variation is much smaller than the outer bounds in the other dimensions. Besides, according to Proposition \ref{th1}, we use the tighter bound $\tilde{\gamma}_{\ell,2}=\gamma_{\ell}/z_{\ell,2}$ for the optimization for MTL classification (we use
	the data of the simulated room temperature to infer the MTL
	formula and the case for the room humidity ratio can be done
	in a similar manner), and by maximizing $z_{\ell,2}$ thus minimizing $\tilde{\gamma}_{\ell,2}$, we can obtain the tightest bound $\tilde{\gamma}^{\ast}_{\ell,2}=\gamma_{\ell}/z^{\ast}_{\ell,2}$. The combined optimization to obtain both $M^{\ast}_{\ell}$ and $z^{\ast}_{\ell,2}$ is as follows:
	
	\begin{align}
	\begin{split}
	&\rm{min}. -z_{\ell,2}^2\\	  
	\rm{s.t.} ~&  M\succ 0,
	A_{\ell}^{T}M_{\ell}+M_{\ell}A_{\ell}\prec 0,\\
	& e_3^{T}Me_3\leq \eta_3, e_4^{T}Me_4\leq \eta_4,\\
	& e_2^{T}Me_2\geq \eta_2, M_{\ell}-z_{\ell,1}^2a_{1,1}^{T}Ma_{1,1}\succeq 0.
	\end{split}
	\end{align}
	where $A_{\ell}$ is the state (or system) matrix in location $\ell$, $e_2=[0,1,0,0]^T, e_3=[0,0,1,0]^T, e_4=[0,0,0,1]^T$, $\eta_2=30$, $\eta_3=\eta_4=10^{-7}$ ($\eta_2$, $\eta_3$ and $\eta_4$ are tuned manually for covering the reset initial sets $\mathcal{L}^1_1\times\mathcal{X}^1_1$ and $\mathcal{L}^1_2\times\mathcal{X}^1_2$).
	
	The optimal solution is computed as $z^{\ast}_{\ell,2}=30$. Based on the two simulated trajectories, we construct
	a timed abstraction (timed automaton) as shown in Fig. \ref{fig_ta_exp} (for details of constructing the timed automaton, see the content of timed abstraction in \cite{YiCDC}). All the events are unobservable except $\psi$ which represents the door opening. We construct a basic observer as in Fig. \ref{fig_obs} (for details of designing the basic observer, see \cite{YiCDC}), where the two occupancy states are never distinguished to the end of the simulation time.
	
	Next we infer an MTL formula that classifies the time-robust tube segments corresponding to the basic observer's states. The observer's initial state $s^1$ contains $(1,1)[0,0]$ and $(2,1)[0,0]$. We first classify the time-robust tube segment $R_{tube}(1,1,[t,t])$ and $R_{tube}(2,1,[t,t])$ ($t\in[0,13]$) but does not find any MTL formula that can achieve perfect classification. Then we move on to state $s^2$ which contains $(1,1)[13,13]$, $(2,1)[13,13]$ and $(2,2)[-6.5,0]$. We find the following formula that perfect clasifies $R_{tube}(1,1,[t,t])$, $R_{tube}(2,1,[t,t])$ and $R_{tube}(2,2,[t-6.5,t])$ ($t\in[0,6.5]$):
	\begin{equation*}
	\phi^{\ast}= \Box_{[1.7717,5]}( x_2\ge 290.6006).
	\end{equation*}
	The optimization takes 36.7 seconds on a laptop with Intel Core i7 and 8GB RAM. 	
	\begin{figure}
		\centering
		\includegraphics[scale=0.45]{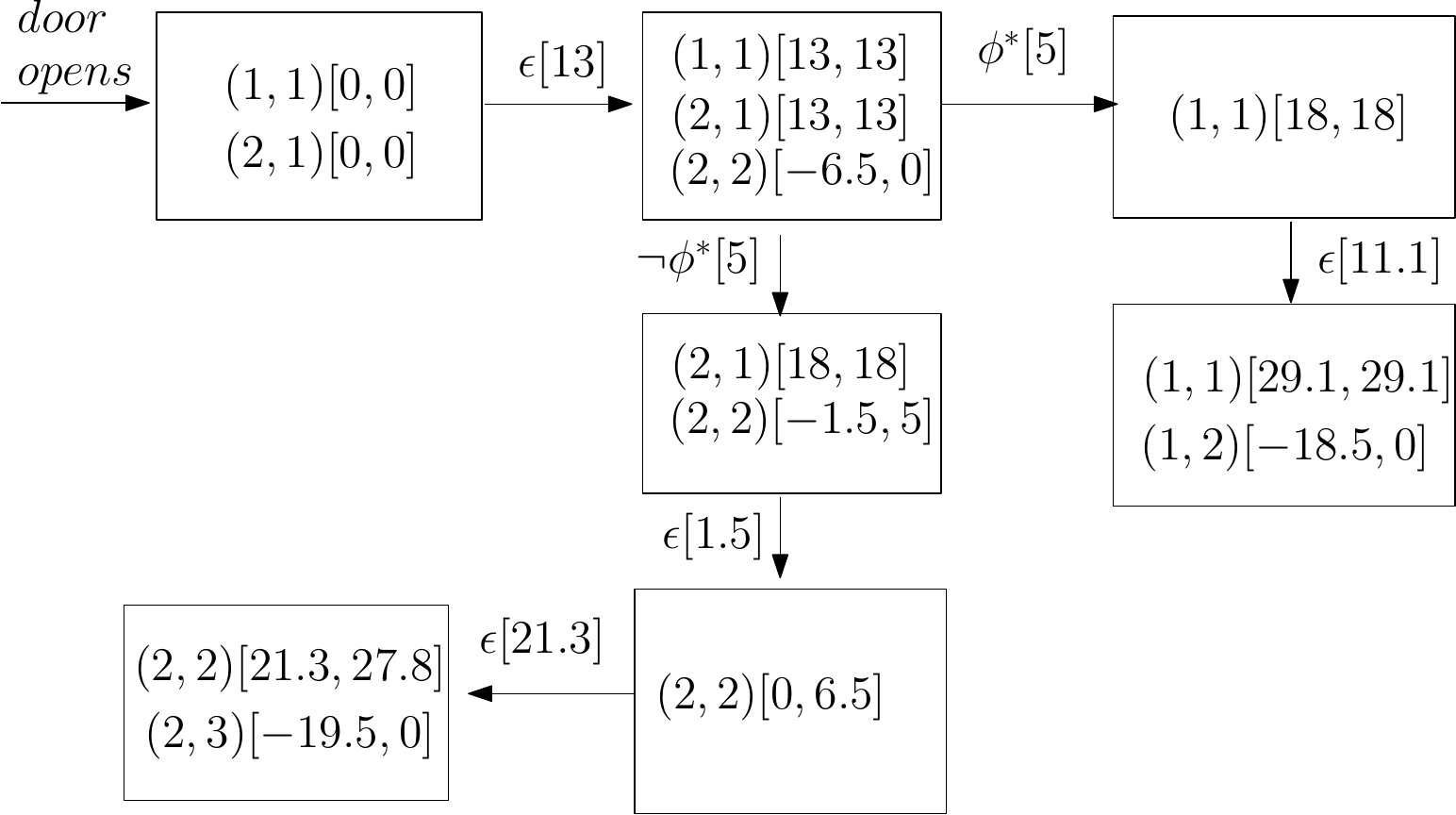}
		\caption{The refined observer shrink the basic observer's states by adding the inferred MTL formula $\phi^{\ast}$. The satisfaction and violation of $\phi$ are modeled as transition labels $\phi^{\ast}[5]$ and $\neg\phi^{\ast}[5]$ respectively.} 
		\label{fig_obs1_exp}
	\end{figure}	
	
	With the inferred MTL formula $\phi^{\ast}$, we construct the refined observer as shown in Fig. \ref{fig_obs1_exp}. It can be seen that once $\phi^{\ast}$ is satisfied, the two cases are distinguished in $18$ seconds. Compared with the basic observer which can never distinguish the two occupancy states, the refinement has achieved the result in 18 seconds by only adding one temperature sensor.
	
	\section{Conclusion}
	We have presented a methodology for occupancy detection of smart building modeled as hybrid systems. It compresses the system states into time-robust tube segments according to trajectory robustness and event occurrence times of the hybrid system, which account for both spatial and temporal uncertainties. Besides occupancy detection, the same methodology can be used in much broader applications such as fault diagnosis, state estimation, etc.
	\section*{Acknowledgment}
	The authors would like to thank Charles C. Okaeme and Dr. Sandipan Mishra for introducing us to the smart building testbed, and Sayan Saha for helpful discussions. This research was partially supported by the National Science
	Foundation through grants CNS-1218109, CNS-1550029 and CNS-1618369.
	
	\section*{APPENDIX}
	\textbf{Proof of proposition \ref{prop_space}}:\\
	We use induction to prove Proposition \ref{prop_space}. 
	
	(i) We first prove that Proposition \ref{prop_space} holds for atomic predicate $\mu$. 
	If $\tau<0$, then $\hat{r}_S(\hat{\xi}_{\ell}(\tau,x^0_{\ell}),\mu)=\hat{r}_S(\hat{\xi}_{\ell}(\tau,\tilde{x}^0_{\ell}),\mu)$ as they both equal to $-\infty$, $\hat{r}_W(\hat{\xi}_{\ell}(\tau,x^0_{\ell}),\mu)=\hat{r}_W(\hat{\xi}_{\ell}(\tau,\tilde{x}^0_{\ell}),\mu)$ as they both equal to $\infty$. Therefore, Proposition \ref{prop_space} trivially holds for $\mu$ if $\tau<0$. If $\tau\ge0$, according to (\ref{semantics}), $\hat{r}_S(\hat{\xi}_{\ell}(\tau,x^0_{\ell}),\mu)=\hat{r}_W(\hat{\xi}_{\ell}(\tau,x^0_{\ell}),\mu)=r(\xi_{\ell}(\tau,x^0_{\ell}),\mu)$, so we only need to prove $r(\hat{\xi}_{\ell}(\tau,x^0_\ell),\mu)-\hat{\gamma}_{\ell}\le r(\hat{\xi}_{\ell}(\tilde{\tau},\tilde{x}^0_\ell),\mu)\le                                                   r(\hat{\xi}_{\ell}(\tau,x^0_\ell),\mu)+\hat{\gamma}_{\ell}$.

	As the metric $d$ satisfies the triangle inequality, we have
	\begin{align}
	\begin{split}
	&d(\xi_{\ell}(\tau,x_\ell^0),y)-d(\xi_{\ell}(\tau,x_\ell^0),\xi_{\ell}(\tau,\tilde{x}_\ell^0)) \le d(\xi_{\ell}(\tau,\tilde{x}_\ell^0),y)\le   \\& d(\xi_{\ell}(\tau,x_\ell^0),y)+d(\xi_{\ell}(\tau,x_\ell^0),\xi_{\ell}(\tau,\tilde{x}_\ell^0)), \forall y\in\mathcal{X}, \tau\in[0,T].
	\end{split}                   
	\label{tri0}                                                      
	\end{align}         
	As $[\big(\xi_\ell(\tau,x^0_\ell)-\xi_\ell(\tau,\tilde{x}^0_\ell)\big)^TM_\ell\big(\xi_\ell(\tau,x^{0}_\ell)-\xi_\ell(\tau,\tilde{x}^0_\ell)\big)]^{\frac{1}{2}}\le\gamma_{\ell}$, we have $d(\xi_{\ell}(\tau,x_\ell^0),\xi_{\ell}(\tau,\tilde{x}_\ell^0))=\norm{\xi_{\ell}(\tau,x_\ell^0)-\xi_{\ell}(\tau,\tilde{x}_\ell^0)}\le\gamma_{\ell}\norm{M_\ell^{-\frac{1}{2}}}=\hat{\gamma}_{\ell}$, thus we have 
	\begin{align}
	&d(\xi_{\ell}(\tau,x_\ell^0),y)-\hat{\gamma}_{\ell} \le d(\xi_{\ell}(\tau,\tilde{x}_\ell^0),y)\le d(\xi_{\ell}(\tau,x_\ell^0),y)+\hat{\gamma}_{\ell}.
	\label{tri}                                        
	\end{align}
	
	1) $\xi_{\ell}(\tau,x_\ell^0)\in \mathcal{O}(\pi)$, and $B_{\ell}(\xi_{\ell}(\tau,x_\ell^0),\gamma_{\ell})\subset\mathcal{O}(\pi)$. In this case, for any $\xi_{\ell}(\tau,\tilde{x}_\ell^0 )\in B_{\ell}(\xi_{\ell}(\tau,x_\ell^0),\gamma_{\ell})$, $r(\xi_{\ell}(\tau,\tilde{x}_\ell^0 ), \pi)=\mbox{inf}\{d(\xi_{\ell}(\tau,\tilde{x}_\ell^0),y)|y \in \mathcal{X}\backslash\mathcal{O}(\pi)\}$.
	From (\ref{tri}), $r(\xi_{\ell}(\tau,\tilde{x}_\ell^0 ), \pi)\ge\mbox{inf}\{d(\xi_{\ell}(\tau,x_\ell^0),y)-\hat{\gamma}_{\ell}|y \in \mathcal{X}\backslash\mathcal{O}(\pi)\}=\mbox{inf}\{d(\xi_{\ell}(\tau,x_\ell^0),y)|y \in \mathcal{X}\backslash\mathcal{O}(\pi)\}-\hat{\gamma}_{\ell}=r(\xi_{\ell}(\tau,x_\ell^0), \pi)-\hat{\gamma}_{\ell}$.
	
	2) $\xi_{\ell}(\tau,x_\ell^0)\notin \mathcal{O}(\pi)$, and $B_{\ell}(\xi_{\ell}(\tau,x_\ell^0),\gamma_{\ell})\subset\mathcal{X}\backslash\mathcal{O}(\pi)$. In this case, for any $\xi_{\ell}(\tau,\tilde{x}_\ell^0)\in B_{\ell}(\xi_{\ell}(\tau,x_\ell^0),\gamma_{\ell})$, $r(\xi_{\ell}(\tau,\tilde{x}_\ell^0), \pi)=-\mbox{inf}\{d(\xi_{\ell}(\tau,\tilde{x}_\ell^0),y)|y \in cl(\mathcal{O}(\pi))\}$.
	From (\ref{tri}), $r(\xi_{\ell}(\tau, \pi), \tau)\ge-\mbox{inf}\{d(\xi_{\ell}(\tau,x_\ell^0),y)+\hat{\gamma}_{\ell}|y \in cl(\mathcal{O}(\pi))\}=r(\xi_{\ell}(\tau,x_\ell^0), \pi)-\hat{\gamma}_{\ell}$.

	3) $\xi_{\ell}(\tau,x_\ell^0)\in \mathcal{O}(\pi)$, but $B_{\ell}(\xi_{\ell}(\tau,x_\ell^0),\gamma_{\ell})\not\subset\mathcal{O}(\pi)$. In this case, we have
	\begin{align} \nonumber
	\begin{split}
	&r(\xi_{\ell}(\tau,\tilde{x}_\ell^0), \pi)\ge\min\limits_{\xi_{\ell}(\tau,\tilde{x}_\ell^0)\in B_{\ell}(\xi_{\ell}(\tau,x_\ell^0),\gamma_{\ell})}r(\xi_{\ell}(\tau,\tilde{x}_\ell^0), \pi)\\
	&~~~~~~~~~~~~~~~~~~~~~~=\min\{X_1, X_2\}, \\
	& \mbox{where}\\
	&X_1=-\max_{\substack{\xi_{\ell}(\tau,\tilde{x}_\ell^0)\in B_{\ell}(\xi_{\ell}(\tau,x_\ell^0),\gamma_{\ell}),\\\xi_{\ell}(\tau,\tilde{x}_\ell^0)\notin \mathcal{O}(\pi)}}\mbox{inf}\{d(\xi_{\ell}(\tau,\tilde{x}_\ell^0),y)|y \in cl(\mathcal{O}(\pi))\},\\
	&X_2=\min_{\substack{\xi_{\ell}(\tau,\tilde{x}_\ell^0)\in B_{\ell}(\xi_{\ell}(\tau,x_\ell^0),\gamma_{\ell}),\\\xi_{\ell}(\tau,\tilde{x}_\ell^0)\in \mathcal{O}(\pi)}}\mbox{inf}\{d(\xi_{\ell}(\tau,\tilde{x}_\ell),y)|y \in \mathcal{X}\backslash\mathcal{O}(\pi)\}.
	\end{split}
	\end{align}      
	
	As $d(\xi_{\ell}(\tau,\tilde{x}_\ell^0),y)\ge0$, so $X_1\le0, X_2\ge0$, $\min\{X_1, X_2\}=X_1$. For any $\xi_{\ell}(\tau,\tilde{x}_\ell^0)\in B_{\ell}(\xi_{\ell}(\tau,x_\ell^0),\gamma_{\ell})$ and $\xi_{\ell}(\tau,\tilde{x}_\ell^0)\notin \mathcal{O}(\pi)$, there exists $z_c\in B_{\ell}(\xi_{\ell}(\tau,x_\ell^0),\gamma_{\ell})$ and $z_c\in \partial(\mathcal{O}(\pi))$ such that $\xi_{\ell}(\tau,\tilde{x}_\ell^0), z_c$ and $\xi_{\ell}(\tau,x_\ell^0)$ are collinear, i.e. $d(\xi_{\ell}(\tau,x_\ell^0),z_c)+d(z_c,\xi_{\ell}(\tau,\tilde{x}_\ell^0))=d(\xi_{\ell}(\tau,x_\ell^0),\xi_{\ell}(\tau,\tilde{x}_\ell^0))\le\hat{\gamma}_{\ell}$. Therefore, as $r(\xi_{\ell}(\tau,x_\ell^0), \pi)=\mbox{inf}\{d(\xi_{\ell}(\tau,x_\ell^0),y)|y \in \mathcal{X}\backslash\mathcal{O}(\pi)\}\le d(\xi_{\ell}(\tau,x_\ell^0),z_c)$ and $\mbox{inf}\{d(\xi_{\ell}(\tau,\tilde{x}_\ell^0),y)|y \in cl(\mathcal{O}(\pi))\}\le d(\xi_{\ell}(\tau,\tilde{x}_\ell^0),z_c)$, we have $\mbox{inf}\{d(\xi_{\ell}(\tau,\tilde{x}_\ell^0),y)|y \in cl(\mathcal{O}(\pi))\}+r(\xi_{\ell}(\tau,x_\ell^0), \pi)\le\hat{\gamma}_{\ell}$ for any $x\in B_{\ell}(x_\ell^0,\gamma_{\ell})$ and $\xi_{\ell}(\tau,\tilde{x}_\ell^0)\notin \mathcal{O}(\pi)$. So $-X_1+r(\xi_{\ell}(\tau,x_\ell^0), \pi)\le\hat{\gamma}_{\ell}$, i.e. $X_1\ge r(\xi_{\ell}(\tau,x_\ell^0), \pi)-\hat{\gamma}_{\ell}$. Therefore, $r(\xi_{\ell}(\tau,\tilde{x}_\ell^0), \pi)\ge\min\{X_1, X_2\}=X_1\ge r(\xi_{\ell}(\tau,x_\ell^0), \pi)-\hat{\gamma}_{\ell}$.
	
	4) $\xi_{\ell}(\tau,x_\ell^0)\notin \mathcal{O}(\pi)$, but  $B_{\ell}(\xi_{\ell}(\tau,x_\ell^0),\gamma_{\ell})\not\subset\mathcal{X}\backslash\mathcal{O}(\pi)$. In this case, $r(\xi_{\ell}(\tau,x_\ell^0), \pi)=-\mbox{inf}\{d(\xi_{\ell}(\tau,x_\ell^0),y)|y \in cl(\mathcal{O}(\pi))\}$. For any $\xi_{\ell}(\tau,\tilde{x}_\ell^0)\in B_{\ell}(\xi_{\ell}(\tau,x_\ell^0),\gamma_{\ell})$ and $\xi_{\ell}(\tau,\tilde{x}_\ell^0)\notin\mathcal{O}(\pi)$, $r(\xi_{\ell}(\tau,x_\ell^0), \pi)=-\mbox{inf}\{d(\xi_{\ell}(\tau,x_\ell^0),y)|y \in cl(\mathcal{O}(\pi))\}\ge-\mbox{inf}\{d(\xi_{\ell}(\tau,x_\ell^0),y)+\hat{\gamma}_{\ell}|y \in cl(\mathcal{O}(\pi))\}=r(\xi_{\ell}(\tau,x_\ell^0), \pi)-\hat{\gamma}_{\ell}$. Therefore, $r(\xi_{\ell}(\tau,\tilde{x}_\ell^0 ), \pi)\ge\min\{X_1, X_2\}=X_1\ge r(\xi_{\ell}(\tau,x_\ell^0), \pi)-\hat{\gamma}_{\ell}$.
	
	(ii) We assume that Proposition \ref{prop_space} holds for $\phi$ and prove Proposition \ref{prop_space} holds for $\lnot\phi$.                                                            
	
	If Proposition \ref{prop_space} holds for $\phi$, then as $\hat{r}_W(\hat{\xi}_{\ell}(\tau,x_\ell^0), \phi)=-\hat{r}_S(\hat{\xi}_{\ell}(\tau,x_\ell^0), \lnot\phi)$, we have $-\hat{r}_S(\hat{\xi}_{\ell}(\tau,x_\ell^0 ), \lnot\phi)-\hat{\gamma}_{\ell}\le -\hat{r}_S(\hat{\xi}_{\ell}(\tau,\tilde{x}_\ell^0), \lnot\phi)\le -\hat{r}_S(\hat{\xi}_{\ell}(\tau,x_\ell^0), \lnot\phi)+\hat{\gamma}_{\ell}$, thus $\hat{r}_S(\hat{\xi}_{\ell}(\tau,x_\ell^0), \lnot\phi)-\hat{\gamma}_{\ell}\le \hat{r}_S(\hat{\xi}_{\ell}(\tau,\tilde{x}_\ell^0), \lnot\phi)\le\hat{r}_S(\hat{\xi}_{\ell}(\tau,x_\ell^0), \lnot\phi)+\hat{\gamma}_{\ell}$. Similarly, as $\hat{r}_S(\hat{\xi}_{\ell}(\tau,x_\ell^0), \phi)=-\hat{r}_W(\hat{\xi}_{\ell}(\tau,x_\ell^0), \lnot\phi)$, we have $-\hat{r}_W(\hat{\xi}_{\ell}(\tau,x_\ell^0 ), \lnot\phi)-\hat{\gamma}_{\ell}\le -\hat{r}_W(\hat{\xi}_{\ell}(\tau,\tilde{x}_\ell^0), \lnot\phi)\le -\hat{r}_W(\hat{\xi}_{\ell}(\tau,x_\ell^0), \lnot\phi)+\hat{\gamma}_{\ell}$, thus $\hat{r}_W(\hat{\xi}_{\ell}(\tau,x_\ell^0), \lnot\phi)-\hat{\gamma}_{\ell}\le \hat{r}_W(\hat{\xi}_{\ell}(\tau,\tilde{x}_\ell^0), \lnot\phi)\le\hat{r}_W(\hat{\xi}_{\ell}(\tau,x_\ell^0), \lnot\phi)+\hat{\gamma}_{\ell}$.	
	
	(iii) We assume that Proposition \ref{prop_space} holds for $\phi_1,\phi_2$ and prove Proposition \ref{prop_space} holds for $\phi_1\wedge\phi_2$. 
	
	If Proposition \ref{prop_space}  holds for $\phi_1$ and $\phi_2$, then $\hat{r}_S(\hat{\xi}_{\ell}(\tau,x_\ell^0 ), \phi_1)-\hat{\gamma}_{\ell}\le \hat{r}_S(\hat{\xi}_{\ell}(\tau,\tilde{x}_\ell^0), \phi_1)\le \hat{r}_S(\hat{\xi}_{\ell}(\tau,x_\ell^0 ), \phi_1)+\hat{\gamma}_{\ell}$, $\hat{r}_S(\hat{\xi}_{\ell}(\tau,x_\ell^0 ), \phi_2)-\hat{\gamma}_{\ell}\le \hat{r}_S(\hat{\xi}_{\ell}(\tau,\tilde{x}_\ell^0), \phi_2)\le \hat{r}_S(\hat{\xi}_{\ell}(\tau,x_\ell^0 ), \phi_2)+\hat{\gamma}_{\ell}$. As $\hat{r}_S(\hat{\xi}_{\ell}(\tau,x_\ell^0 ), \phi_1\wedge\phi_2)=\min(\hat{r}_S(\hat{\xi}_{\ell}(\tau,x_\ell^0 ), \phi_1),\hat{r}_S(\hat{\xi}_{\ell}(\tau,x_\ell^0 ), \phi_2))$, we have 
	\begin{align}\nonumber
	\begin{split}
	&\min(\hat{r}_S(\hat{\xi}_{\ell}(\tau,x_\ell^0 ), \phi_1),\hat{r}_S(\hat{\xi}_{\ell}(\tau,x_\ell^0 ), \phi_2))-\hat{\gamma}_{\ell}\le \hat{r}_S(\hat{\xi}_{\ell}(\tau,\tilde{x}_\ell^0), \\&\phi_1\wedge\phi_2)\le \min(\hat{r}_S(\hat{\xi}_{\ell}(\tau,x_\ell^0 ), \phi_1),\hat{r}_S(\hat{\xi}_{\ell}(\tau,x_\ell^0 ), \phi_2))+\hat{\gamma}_{\ell},
	\end{split}
	\end{align}
	therefore $\hat{r}_S(\hat{\xi}_{\ell}(\tau,x_\ell^0 ), \phi_1\wedge\phi_2)-\hat{\gamma}_{\ell}\le \hat{r}_S(\hat{\xi}_{\ell}(\tau,\tilde{x}_\ell^0), \phi_1\wedge\phi_2)\le \hat{r}_S(\hat{\xi}_{\ell}(\tau,x_\ell^0 ), \phi_1\wedge\phi_2)+\hat{\gamma}_{\ell}$. 
	
	Similarly, it can be proved that if Proposition \ref{prop_space} holds for $\phi_1$ and $\phi_2$, then $\hat{r}_W(\hat{\xi}_{\ell}(\tau,x_\ell^0 ), \phi_1\wedge\phi_2)-\hat{\gamma}_{\ell}\le \hat{r}_W(\hat{\xi}_{\ell}(\tau,\tilde{x}_\ell^0), \phi_1\wedge\phi_2)\le\hat{r}_W(\hat{\xi}_{\ell}(\tau,x_\ell^0 ), \phi_1\wedge\phi_2)+\hat{\gamma}_{\ell}$.                                                   
	                                                                                                  
	(iv) We assume that Proposition \ref{prop_space} holds for $\phi$ and prove Proposition \ref{prop_space} holds for $F_{I}\phi$.                                          	
	
	As $\hat{r}_S(\hat{\xi}_{\ell}(\tau,x_\ell^0), F_{\mathcal{I}}\phi)=\displaystyle\max_{\tau'\in (t+\mathcal{I})}\hat{r}_S(\hat{\xi}_{\ell}(\tau',x_\ell^0), \phi)$, $\hat{r}_W(\hat{\xi}_{\ell}(\tau,x_\ell^0), F_{\mathcal{I}}\phi)=\displaystyle\max_{\tau'\in (t+\mathcal{I})}\hat{r}_W(\hat{\xi}_{\ell}(\tau',x_\ell^0), \phi)$, if Proposition \ref{prop_space} holds for $\phi$, then for any $\tau'\in (t+\mathcal{I})$, $\hat{r}_S(\hat{\xi}_{\ell}(\tau',x_\ell^0), \phi)-\hat{\gamma}_{\ell}\le \hat{r}_S(\hat{\xi}_{\ell}(\tau',\tilde{x}_{\ell}), \phi)\le\hat{r}_S(\hat{\xi}_{\ell}(\tau',x_\ell^0), \phi)+\hat{\gamma}_{\ell}$, $\hat{r}_W(\hat{\xi}_{\ell}(\tau',x_\ell^0), \phi)-\hat{\gamma}_{\ell}\le \hat{r}_W(\hat{\xi}_{\ell}(\tau',\tilde{x}_{\ell}), \phi)\le\hat{r}_W(\hat{\xi}_{\ell}(\tau',x_\ell^0), \phi)+\hat{\gamma}_{\ell}$. So we have 
	\begin{align}\nonumber            
	\begin{split}
	&\max_{\tau'\in (t+\mathcal{I})}\hat{r}_S(\hat{\xi}_{\ell}(\tau',x_\ell^0), \phi)-\hat{\gamma}_{\ell}\le\max_{\tau'\in (\tau+\mathcal{I})}\hat{r}_S(\hat{\xi}_{\ell}(\tau',\tilde{x}_{\ell}), \phi)\le\\
	&\max_{\tau'\in (t+\mathcal{I})}\hat{r}_S(\hat{\xi}_{\ell}(\tau',x_\ell^0), \phi),\\
	&\max_{\tau'\in (t+\mathcal{I})}\hat{r}_W(\hat{\xi}_{\ell}(\tau',x_\ell^0), \phi)-\hat{\gamma}_{\ell}\le\max_{\tau'\in (\tau+\mathcal{I})}\hat{r}_W(\hat{\xi}_{\ell}(\tau',\tilde{x}_{\ell}), \phi)\le\\
	&\max_{\tau'\in (t+\mathcal{I})}\hat{r}_W(\hat{\xi}_{\ell}(\tau',x_\ell^0), \phi).	  
	\end{split}
	\end{align}
	Thus Proposition \ref{prop_space} holds for $F_{\mathcal{I}}\phi$.
	Similarly, it can be proved that if Proposition \ref{prop_space} holds for $\phi$, then Proposition \ref{prop_space} holds for $G_{\mathcal{I}}\phi$.  
	Therefore, it is proved that Proposition \ref{prop_space} holds for any $\phi$.

	\textbf{Proof of Proposition \ref{th1}}:\\
	The proof of Proposition \ref{th1} is similar with that of Proposition \ref{prop_space} except the case for the atomic predicate $\mu$ when $\tau\ge0$.	
	
	We denote $\Pi_j:\mathbb{X}\rightarrow\mathbb{X}_j$ as a projection map that maps every state $x\in\mathbb{X}$ to its value at the $i$th dimension, i.e. $\Pi_j(x)=x_j=e_jx$, where $e_j$ is a canonical unit vector. From (\ref{tri0}), as $[\big(\xi_\ell(\tau,x^0_\ell)-\xi_\ell(\tau,\tilde{x}^0_\ell)\big)^Tz_{\ell,j}^2e_j^{T}e_j\big(\xi_\ell(\tau,x^{0}_\ell)-\xi_\ell(\tau,\tilde{x}^0_\ell)\big)]^{\frac{1}{2}}\le[\big(\xi_\ell(\tau,x^0_\ell)-\xi_\ell(\tau,\tilde{x}^0_\ell)\big)^TM_\ell\big(\xi_\ell(\tau,x^{0}_\ell)-\xi_\ell(\tau,\tilde{x}^0_\ell)\big)]^{\frac{1}{2}}\le\gamma_{\ell}$, we have $d(\Pi_j(\xi_{\ell}(\tau,x_\ell^0)),\Pi_j(\xi_{\ell}(\tau,\tilde{x}_\ell^0)))=[\big(\xi_\ell(\tau,x^0_\ell)-\xi_\ell(\tau,\tilde{x}^0_\ell)\big)^Te_j^{T}e_j\big(\xi_\ell(\tau,x^{0}_\ell)-\xi_\ell(\tau,\tilde{x}^0_\ell)\big)]^{\frac{1}{2}}\le\gamma_{\ell}/z_{\ell,j}$, thus we have 
	\begin{align}
	\begin{split}    
	& d(\Pi_j(\xi_{\ell}(\tau,x_\ell^0)),\Pi_j(y))-\tilde{\gamma}_{\ell,j} \le d(\Pi_j(\xi_{\ell}(\tau,\tilde{x}_\ell^0)),\Pi_j(y))\\&\le d(\Pi_j(\xi_{\ell}(\tau,x_\ell^0)),\Pi_j(y))+\tilde{\gamma}_{\ell,j}.
	\end{split}                               
	\label{tri2}            
	\end{align}
	The remaining proof is similar to the proof of Proposition \ref{prop_space}, replacing $d(x,y)$ by $d(\Pi_j(x),\Pi_j(y))$ and $\hat{\gamma_{\ell}}$ by $\tilde{\gamma}_{\ell,j}$.
	
	\textbf{Proof of Proposition \ref{sol_th}}:\\
	For $c_i=1$, according to Proposition \ref{prop_space}, for any $\big(\tau_i,\hat{\xi}_{\ell^{m_i}_{k_i}}(\tau_i,\tilde{x}^0_{\ell^{m_i}_{k_i}})\big)\in R_{tube}(k_i,m_i,[t+\bar{a}_i,t+\bar{b}_i])$, we have $\hat{r}_W(\hat{\xi}_{\ell^{m_i}_{k_i}}(\tau_i,x^0_{\ell^{m_i}_{k_i}}),\phi)-\hat{\gamma}_{\ell^{m_i}_{k_i}}\le \hat{r}_W(\hat{\xi}_{\ell^{m_i}_{k_i}}(\tau_i,\tilde{x}^0_{\ell^{m_i}_{k_i}}),\phi)$. If $MG(k_i,m_i,\bar{a}_i,\bar{b}_i,\phi,c_i)=\min\limits_{\tau_i\in t+ [\bar{a}_i,\bar{b}_i]}\hat{r}_W(\hat{\xi}_{\ell^{m_i}_{k_i}}(\tau_i,x^0_{\ell^{m_i}_{k_i}}),\phi)-\hat{\gamma}_{\ell^{m_i}_{k_i}}>0$, then for any $\tau_i\in t+ [\bar{a}_i,\bar{b}_i]$, $\hat{r}_W(\hat{\xi}_{\ell^{m_i}_{k_i}}(\tau_i,\tilde{x}^0_{\ell^{m_i}_{k_i}}),\phi)\ge \hat{r}_W(\hat{\xi}_{\ell^{m_i}_{k_i}}(\tau_i,x^0_{\ell^{m_i}_{k_i}}),\phi)-\hat{\gamma}_{\ell^{m_i}_{k_i}}>0$.
	
	For $c_i=-1$, according to Proposition \ref{prop_space}, for any $\big(\tau_i,\hat{\xi}_{\ell^{m_i}_{k_i}}(\tau_i,\tilde{x}^0_{\ell^{m_i}_{k_i}})\big)\in R_{tube}(k_i,m_i,[t+\bar{a}_i,t+\bar{b}_i])$, we have $\hat{r}_W(\hat{\xi}_{\ell^{m_i}_{k_i}}(\tau_i,x^0_{\ell^{m_i}_{k_i}}),\lnot\phi)-\hat{\gamma}_{\ell^{m_i}_{k_i}}\le \hat{r}_W(\hat{\xi}_{\ell^{m_i}_{k_i}}(\tau_i,\tilde{x}^0_{\ell^{m_i}_{k_i}}),\lnot\phi)$. If $MG(k_i,m_i,\bar{a}_i,\bar{b}_i,\lnot\phi,c_i)=\min\limits_{\tau_i\in t+ [\bar{a}_i,\bar{b}_i]}\hat{r}_W(\hat{\xi}_{\ell^{m_i}_{k_i}}(\tau_i,x^0_{\ell^{m_i}_{k_i}}),\lnot\phi)-\hat{\gamma}_{\ell^{m_i}_{k_i}}>0$, then for any $\tau_i\in t+ [\bar{a}_i,\bar{b}_i]$, $\hat{r}_W(\hat{\xi}_{\ell^{m_i}_{k_i}}(\tau_i,\tilde{x}^0_{\ell^{m_i}_{k_i}}),\lnot\phi)\ge \hat{r}_W(\hat{\xi}_{\ell^{m_i}_{k_i}}(\tau_i,x^0_{\ell^{m_i}_{k_i}}),\lnot\phi)-\hat{\gamma}_{\ell^{m_i}_{k_i}}>0$. 

	\bibliographystyle{IEEEtran}
	\bibliography{yd}
	
\end{document}